# Networked Collaborative Sensing using Multi-domain Measurements: Architectures, Performance Limits and Algorithms

Yihua Ma, Shuqiang Xia, Chen Bai, Zhongbin Wang, and Songqian Li

*Abstract*—As a promising 6G technology, integrated sensing and communication (ISAC) gains growing interest. ISAC provides integration gain via sharing spectrum, hardware, and software. However, concerns exist regarding its sensing performance when compared to the dedicated radar. To address this issue, the advantages of widely deployed networks should be utilized. This paper proposes networked collaborative sensing (NCS) using multi-domain measurements (MM), including range, Doppler, and two-dimension angles. For the NCS-MM architecture, this paper proposes a novel multi-domain decoupling model and a novel guard band-based protocol. The proposed model simplifies multi-domain derivations and algorithm designs, and the proposed protocol conserves resources and mitigates NCS interference. In terms of performance limits, this paper derives the Cramér-Rao lower bound (CRLB) of position and velocity estimations in NCS-MM. An accumulated single-dimension channel model is proposed, which is proven to be equivalent to that of the multi-dimension model. The algorithms of both MM estimation and fusion are proposed. An arbitrary-dimension Newtonized orthogonal matched pursuit (AD-NOMP) is proposed to accurately estimate grid-less MM. The degree-of-freedom (DoF) of MM is analyzed, and a novel DoF-based two-stage weighted least squares (TSWLS) is proposed to reduce complexity without DoF loss. The numerical results show that the proposed algorithms approach their performance limits.

*Index Terms*—Integrated sensing and communication (ISAC), networked collaborative sensing, multi-domain decoupling model, multi-dimension frequency estimation, two-stage weighted least squares (TSWLS).

## I. INTRODUCTION

OWING to the rapid development of wireless communications, integrated sensing and communication (ISAC) [1]–[2] has become a very promising 6G technology. It empowers future communication systems to not only transmit data but also extract valuable information from the physical world. ISAC has gained growing interest from both academic and industrial fields, leading to a rise in research efforts [4]–[7]. Meanwhile, the ISAC related global standardization has begun [8], [9].

ISAC schemes are categorized into three types: coexistence, cooperation, and joint design. The coexistence scheme uses overlapped resources [10]. The cooperation scheme facilitates interference management through spectrum sharing [11] and beamforming [12]. Cognitive radar [13] is a special spectrum sharing, which dynamically identifies idle resources for transmission. The joint design approach merges the two sub-systems into one. It enables joint optimization, e.g. waveform optimization [14] and power optimization [15]. Additionally, the orthogonal frequency division multiplexing (OFDM) waveform can be flexibly utilized. The unused sub-carriers are filled with optimized data for sensing performance [16], while sensing overheads are reduced via non-uniform sampling [17] and max-aperture radar slicing [18].

Apart from designing a highly integrated system, another crucial issue is to boost the sensing ability of the network and make it competitive with radars. Radars have superiority in terms of self-interference separation, phase continuity, and synthetic aperture. The advantage of ISAC is to use the widely deployed networks to realize networked sensing [1], or collaborative sensing [19]. The networked collaborative sensing (NCS) has many benefits, including target resolution, low-speed target identification, and robustness against target scintillation [20]. Low-cost target monitoring terminals are proposed to realize widespread sensing, and the communication energy is leaked for sensing to increase energy efficiency [21]. A two-phase sensing framework [22] is proposed to measure distances at several nodes and then use maximum-likelihood estimation (MLE) to obtain a single position. Based on this framework, a joint data association, non-line-of-sight (NLOS) mitigation, and clutter suppression algorithm are further proposed [23]. A networking based ISAC hardware testbed [24] shows the performance gain over single-station sensing. Also, a multi-cell cooperative sensing method is proposed to realized an edge intelligence oriented ISAC [25]. However, the NCS using multi-domain measurements has not been researched in the ISAC area.

In radars, extensive researches have been conducted on multiple-input and multiple-output (MIMO) radar with widely separated antennas, or distributed MIMO radar. Most works

This work was supported by the National Key Research and Development Program of China (No. 2021YFB2900200). *(Corresponding author: Yihua Ma.)*

The authors are with the State Key Laboratory of Mobile Network and Mobile Multimedia Technology, Shenzhen, China. They are also with the ZTE Corporation, Shenzhen, China. (e-mail: {yihua.ma, xia.shuqiang, chen.bai1, wang.zongbin, li.songqian}@zte.com.cn)

Mentions of supplemental materials and animal/human rights statements can be included here.

Color versions of one or more of the figures in this article are available online at http://ieeexplore.ieee.org

[26]–[31] merge distance measurements for localization. In [26], the Cramér-Rao lower bound (CRLB) of localization is given. Both MLE and the best linear unbiased estimator (BLUE) are proposed [26]. BLUE is low-complex but requires an accurate initial estimation. One-stage least squares (OSLS) [27] is proposed without the need of initial guess. Two-stage weighted least squares (TSWLS) [28] is proposed to obtain more accurate estimations. Variants of TSWLS [31]–[31] are proposed to approach the CRLB. There are several positioning and radar works [32]–[34] that consider joint localization and velocity estimation using multi-domain measurements (MM) of range, Doppler, and angle-of-arrival (AoA). However, the radar works among them either require many stages of weighted least squares (WLS) [33] or use extra angle rates, which is not easy to obtain in practice [34]. Also, the modelling of MM in ISAC is also important. A unified tensor framework [35] has been proposed for ISAC, but the coupling of 2D-AoA has not been considered yet.

This paper proposes NCS using MM of range, Doppler, and 2D-AoA for ISAC. Similar cases have been considered in some radar work [33], [34], but there are still unresolved issues. First, the classical 2D-AoA model cannot obtain a diagonal covariance matrix of MM as the elevation and azimuth angle are coupled. Second, the existing CRLB of NCS-MM is based on a known covariance matrix of MM, which requires further derivations. Third, the MM estimation step is not considered, and fusion algorithm algorithms construct $\mathcal{O}(IJ)$ redundant equations, where $I$ and $J$ are the number of transmitters and receivers. The redundancy leads to computational inefficiency, especially when the NCS scale is large. To solve these problems, this paper proposes a multi-domain decoupling model for NCS-MM in the first step, which simplifies the CRLB derivations and MM estimation algorithms. The CRLB of MM is derived using this decoupling model, and then the CRLB of position and velocity are derived. In terms of algorithms, Newtonized orthogonal matched pursuit (NOMP) is extended to arbitrary dimension NOMP (AD-NOMP) basing on the proposed decoupling model. The degree of freedom (DoF)-based TSWLS is proposed to construct equations according to the DoF of MM in both full-duplex (HD)-NCS and half-duplex (HD) NCS, which removes the redundancy of equations. Apart from the computation efficiency, the interference is also an important issue in NCS, which is more complex than that in radar. Some protocol designs are proposed to solve the interference issues.

The main contributions of this paper are summarized as follows:

- This paper proposes a multi-domain decoupling model and a guard period (GP)-based protocol design for NCS-MM. The proposed decoupling model makes the CRLB derivation of MM in each domain possible and allows unified estimation processing in different domains. The proposed protocol uses the empty GP in communications for sensing with flexible timing strategies to avoid NCS interference with propagation delays. A novel cyclic shifts-based multiplexing method is also proposed.

- This paper derives the CRLB of MM and the CRLB of 3D position and velocity for NCS-MM. Existing works [33], [34] directly obtain the CRLB with a known covariance matrix of MM. However, that matrix is not easy to obtain as the 2D-AoAs are coupled. Based on the proposed multi-domain decoupling model, this paper uses an accumulated single-dimension model to derive the CRLB of MM, which is proven to be equivalent to that of a multi-dimension model. Then, the CRLB of position and velocity is given using the chain rule of Fisher information matrix (FIM) and used to verify algorithms and instruct NCS system design.

- This paper proposes an iterative algorithm of AD-NOMP for MM estimation and a closed-form algorithm of DoF-TSWLS for MM fusion. With decoupling dimensions, AD-NOMP employs a unified process in each dimension to obtain an accurate grid-free estimation of MM. This paper analyzes the DoF of MM and finds that the range, or Doppler, measurements have a DoF of $N_{BS}$ and $N_{BS}-1$ in FD-NCS and HD-NCS, respectively, where $N_{BS}$ is the base station number. DoF-TSWLS reduces the number of equations from $\mathcal{O}(IJ)$ [33], [34] to $\mathcal{O}(N_{BS})$, while the performances are still close to the CRLB.

The rest of the paper is organized as follows: Section II introduces the multi-domain decoupling model and the NCS protocol design. Section III derives the CRLB of MM as well as the CRLB of 3D position and velocity. Section IV proposes AD-NOMP for MM estimations and DoF-TSWLS for joint position and velocity estimation. The numerical results are shown in Section V to verify the proposed schemes. Finally, Section VI briefly concludes this paper. In this paper, $(\cdot)^*$, $(\cdot)^T$, and $(\cdot)^H$ denote the conjugate, transpose, and conjugate transpose of a matrix or vector. $|\cdot|$ denotes absolute value, and $\|\cdot\|$ denotes the L2 norm. $\otimes$ denotes the Kronecker product. $\mathbf{0}_N$ and $\mathbf{1}_N$ represent a all-zero and all-one vector of length $N$. $\mathbf{I}_N$ denotes an $N \times N$ identity matrix. $\mathcal{R}(\cdot)$ is a function of getting real parts.

## II. SYSTEM MODEL AND PROTOCOL DESIGN

As illustrated in Fig. 1, this paper considers two NCS scenarios, including HD-NCS and FD-NCS. In HD-NCS, the cooperative BSs are all half-duplex. In FD-NCS, part or all of the cooperative BSs support full duplex. The line-of-sight paths between the BSs and the targets are used for wireless sensing. Assume that there are $N_{BS}$ BSs and $K$ targets. Among these BSs, there are $I$ transmitting (TX) BSs, which are denoted by BS $i$, where $i = 0, 1, ..., I-1$. $J$ receiving (RX) BSs are represented by BS $j$, where $j = J_S, J_S+1, ..., N_{BS}-1$, and $J_S = N_{BS} - J$. $J_S = I$ in HD-NCS, while $J_S = 0$ in FD-NCS. At the transmitter, the utilization of omnidirectional or wide beams is assumed to avoid the time-consuming beam scanning in NCS. Without extra explanations, the $(i,j)$ pair is sorted as $(i,j) = [(0,J_S), (0,J_S + 1), ..., (0,J + J_S - 1), ..., (I - 1, J + J_S - 1)]$ and denoted as the $l$-th pair with $l = iJ + j - J_S = 0, 1, ..., IJ - 1$. This paper also employs the variables $i_0 = 0, 1, ..., N_{BS} - 1$, and $i_1 = 1, 2, ..., N_{BS} - 1$, in addition to $i$, to make a distinction. The sensing signals are transmitted through $M$ orthogonal frequency-division multiplexing (OFDM) symbols, which have a repetition period of $T$. Each symbol comprises $N$ sub-carriers, and the sub-carrier spacing is represented by $\Delta f$. The carrier frequency is $f_C$. Each BS is equipped with a receiving uniform planar array (UPA) of $L_X L_Y$ half-wavelength spacing antennas, where $L_X$ and $L_Y$ denote the number of antennas in the horizontal and vertical directions, respectively.

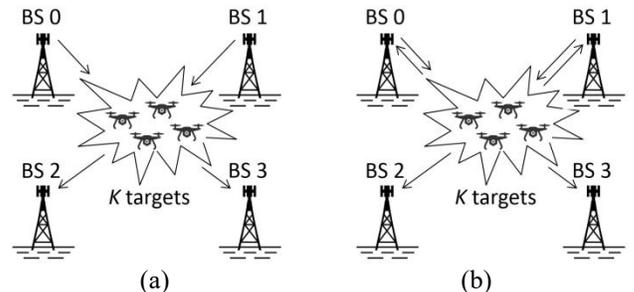

**Fig. 1.** The illustration of (a) HD-NCS and (b) FD-NCS system. The number of TX BSs is from 1 to $N_{BS}$, and TX BSs uses the same time-frequency resources.

*A. Multi-domain Decoupling Model*

This paper proposes a multi-domain decoupling model to model MM. The range, Doppler, and 2D AoA are modeled as the decoupling frequencies of a high-order tensor. The model describes the multi-domain channel from the $i$-th BS to the $k$-th target and then scattering to the $j$-th BS. Note the echo signals of different targets will be overlapped, and the MM estimation method is required to separate targets with different MM parameters, which will be introduced in Section IV.

The total range $r_{i,j,k} = d_{i,k} + d_{j,k}$ is carried in the frequency-domain channel with $N$ sub-carriers, where $d_{i_0,k}$ is the distance between BS $i_0$ and target $k$. The normalized frequency-domain channel is

$$\mathbf{h}_{i,j,k}^{freq} = e^{-j2\pi\left(f_c + \left([0,1,...,N-1]^T - \frac{N}{2}\right)\Delta f\right)\frac{r_{i,j,k}}{c_0}}$$
$$\triangleq e^{-j2\pi\left(\frac{r_{i,j,k}\Delta f}{c_0}[0,1,...,N-1]^T + \phi_{i,j,k}\right)} \in \mathbb{C}^{N\times 1}, \quad (1)$$

where $c_0$ is the speed of light, and $\phi_{i,j,k} \triangleq (f_c - N\Delta f/2)r_j/c_0$ is the channel initial phase.

The total radial velocity of $\dot{r}_{i,j,k} = \dot{d}_{i,k} + \dot{d}_{j,k}$ is carried in the channels of $M$ symbols, where $\dot{d}_{i_0,k}$ is the radial velocity of target $k$ from the view of BS $i_0$. This domain is named slow time in radar [37], and the normalized slow-time-domain channel is

$$\mathbf{h}_{i,j,k}^{st} = e^{j2\pi\frac{\dot{r}_{i,j,k}f_cT}{c_0}[0,1,...,M-1]^T} \in \mathbb{C}^{M\times 1}. \quad (2)$$

With Equations (1) and (2), the frequency-domain channel information of $M$ symbols is $\mathbf{H}_{i,j,k}^{tf} = \mathbf{h}_{i,j,k}^{f}\left(\mathbf{h}_{i,j,k}^{t}\right)^T \in \mathbb{C}^{N\times M}$, which can be used to generate the rang-Doppler map [37], [38].

As shown in Fig. 2, the classical 2D-AoA model uses the azimuth angle $\phi_{j,k}$ and the elevation angle $\theta_{j,k}$ in spherical coordinates. Although these two angles are widely used, the corresponding estimation is coupled for the UPA. The coupling of 2D AoAs also makes bound derivation and algorithm design more complex. To solve this problem, this paper proposes to use decoupling 2D AoAs of horizontal angle $\alpha_{j,k}$ and vertical angle $\beta_{j,k}$ [39]. The relationship between the two 2D AoA models is

$$\sin(\theta_{j,k})\cos(\phi_{j,k}) = \cos(\alpha_{j,k}),$$
$$\sin(\theta_{j,k})\sin(\phi_{j,k}) = \cos(\beta_{j,k}). \quad (3)$$

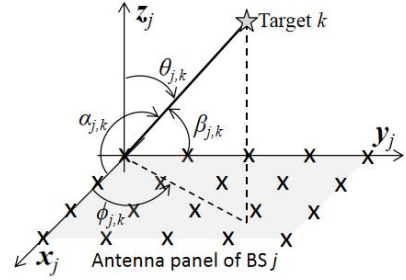

**Fig. 2.** The two types of 2D-AoA models at the BS $j$.

The normalized 2D spatial-domain channel information is

$$\mathbf{H}_{i,j,k}^{space} = \mathbf{a}(\alpha_{j,k})\mathbf{a}(\beta_{j,k})^T \in \mathbb{C}^{L_X\times L_Y}, \quad (4)$$

where $\mathbf{a}(\alpha) = e^{j\pi\cos(\alpha)[0,1,...,L_X-1]^T} \in \mathbb{C}^{L_X\times 1}$. Equation (4) can also be vectorized as

$$\mathbf{h}_{i,j,k}^{ang} = \text{vec}\left(\mathbf{H}_{i,j,k}^{ang}\right) = \mathbf{a}(\beta_{j,k}) \otimes \mathbf{a}(\alpha_{j,k}) \in \mathbb{C}^{L_X L_Y\times 1}, \quad (5)$$

This paper further proposes to directly use the cosine values of $\alpha_{j,k}$ and $\beta_{j,k}$ in the MM model. They can be seen as decoupled frequencies of a tensor, and they can be directly expressed by

$$\cos(\alpha_{j,k}) = \mathbf{x}_j^T \boldsymbol{\rho}_{\mathbf{t}_k,\mathbf{b}_j},$$
$$\cos(\beta_{j,k}) = \mathbf{y}_j^T \boldsymbol{\rho}_{\mathbf{t}_k,\mathbf{b}_j}. \quad (6)$$

where $\mathbf{x}_j$ and $\mathbf{y}_j$ are the normalized vertical and horizontal direction vectors of the antenna panel in BS $j$, and $\boldsymbol{\rho}_{\mathbf{t}_k,\mathbf{b}_j} \triangleq (\mathbf{t}_k - \mathbf{b}_j)/\|\mathbf{t}_k - \mathbf{b}_j\|$ is the normalized direction vector from the position vector of target $k$, to that of the BS $j$.

The channel coefficient is

$$A_{i,j,k} = \sqrt{\frac{P_{T,i}G_{T,i}G_{R,j}\lambda^2\delta_{i,j,k}}{(4\pi)^3 d_{i,k}^2 d_{j,k}^2}}, \quad (7)$$

where $P_{T,i}$, $G_{T,i}$, $G_{R,j}$, $\lambda$ and $\delta_{i,j,k}$ denote the TX power of BS $i$, TX gain of BS $i$, RX gain of BS $j$, wavelength, and the radar cross section (RCS) of target $k$ from the view of the $(i,j)$ pair, respectively. The entire sensing channel from the $i$-th BS to the $j$-th BS is a 4-dimension array, or a 4-order tensor, which is written as

$$\mathbf{h}_{i,j,k} = A_{i,j,k}\mathbf{h}_{i,j,k}^{ang} \otimes \mathbf{h}_{i,j,k}^{st} \otimes \mathbf{h}_{i,j,k}^{freq} \in \mathbb{C}^{NML_XL_Y}, \quad (8)$$

To get a more unified form, Equation (8) is rewritten to

$$\mathbf{h}_{i,j,k} = A_{i,j,k}e^{j\phi_{i,j,k}}\boldsymbol{\Lambda}_{i,j,k}^{DM3} \otimes \boldsymbol{\Lambda}_{i,j,k}^{DM2} \otimes \boldsymbol{\Lambda}_{i,j,k}^{DM1} \otimes \boldsymbol{\Lambda}_{i,j,k}^{DM0}, \quad (9)$$

where

$$\begin{aligned}
\mathbf{\Lambda}_{i,j,k}^{\text{DMa}} &= e^{j2\pi f_{i,j,k}^{\text{DMa}}[0,1,\ldots,N_a-1]^T}, a = 0,1,2,3, \\
f_{i,j,k}^{\text{DM0}} &= 1 - r_{i,j,k}\Delta f/c_0 \in [0,1), N_0 = N, \\
f_{i,j,k}^{\text{DM1}} &= \dot{r}_{i,j,k}f_C T/c_0 \in [-0.5,0.5), N_1 = M, \\
f_{i,j,k}^{\text{DM2}} &= \cos\alpha_{i,j,k}/2 \in [-0.5,0.5), N_2 = L_x, \\
f_{i,j,k}^{\text{DM3}} &= \cos\beta_{i,j,k}/2 \in [-0.5,0.5), N_3 = L_y.
\end{aligned} \qquad (10)$$

In this model, the range of $f_{i,j,k}^{\text{DM0}}$ is different as the range is always non-negative.

*B. NCS Protocol Design*

This paper proposes a protocol design to solve the NCS interference issues. The interference in NCS is categorized to three types. The first is interference from sensing to communication. It can greatly affect uplink communications, as the uplink signal is relatively weak. The second is from communication to sensing. As the sensing echoes are weak, the downlink communication signals can easily conceal them. The last is from sensing to sensing when more than one TX sensing BS is employed. The time-division method can avoid it but consume multi-fold resources. Moreover, it can be easy to make zero-delay sensing signals orthogonal to each other, but the orthogonality is hard to maintain for the echo signals with different delays and Doppler frequencies.

Apart from the NCS interference, the NCS timing is also important to coordinate multiple BSs in NCS. In both uplink and downlink communications, the timing adjustment of UE is to align the TX and RX window according to time advance (TA). Then, the CP of the OFDM symbol only requires to cover the delay spread instead of the maximum delay. For NCS, such an operation is also required to save CP overheads. However, existing timing adjustments are done by UEs, and BSs cannot implement such operations.

To solve these problems, several protocol enhancements are proposed as follows:

***Sensing in GP:*** As in Fig. 3(a), the GP [40] is left blank in conventional communication, which gives time for UE to switch from DL to UL under certain propagation delays. Considering the

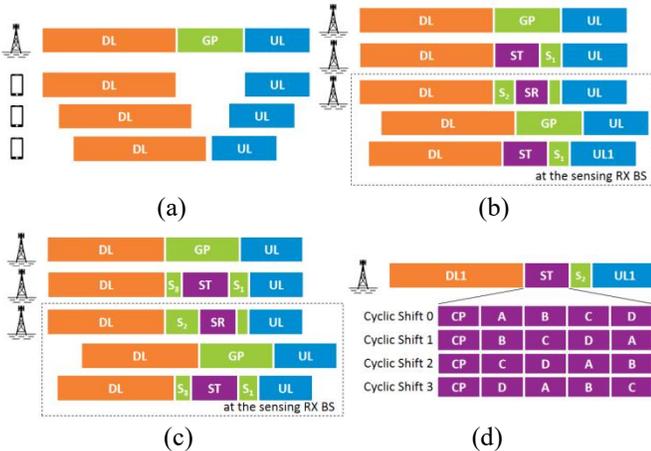

**Fig. 3.** The frame structure of (a) the GP in communication, (b) sensing in GP, (c) cyclic shift multiplexing, and (d) the enhancement of BS timing adjustment.

TA in DL and UL, the length of the GP has to satisfy

$$T_{GP} \geq 2R_{cell}/c_0, \qquad (11)$$

where $R_{cell}$ is the cell radius. This paper proposes to send the sensing signal in the empty GP, which reuses these resources. Apart from the resource utilization improvement, another advantage is to avoid the interference between sensing and communications. There are two gaps marked as $S_1$ and $S_2$ as shown in Fig. 3(b). $S_1$ avoids the sensing signal from interfering with the communications, while $S_2$ prevents the interference of the downlink communications signal from entering the sensing RX window. In normal communication slots, such operations can be hard to implement as the transmission timing is fixed in each slot. As the GP is empty and only used for sensing, it is more flexible to adjust the transmission parameters, including TX/RX timing and OFDM numerology to adapt the sensing requirements.

***BS Timing Adjustment:*** In communications, UEs implement the timing adjustment, and it is hard for BS to adjust the TX or RX window as it needs to serve a lot of UEs. This paper proposes a BS timing adjustment strategy in GP. The receiving BS $j$ pushes back the receiving window by a propagation delay of $\min(d_{i,j}/c_0)$ for all $i$. This operation saves the CP overheads by $\min(d_{i,j}/c_0)$. As in Fig. 3(b), an empty gap $S_2$ has been added to eliminate the interference, and it can be reused for time adjustment when

$$d_{int,j}/c_0 \leq T_{S_2} \leq \min(d_{i,j}/c_0), \qquad (12)$$

where $d_{int,j}$ is the maximum distance between the communication interference BS to the receiving BS. However, these two limitations cannot be simultaneously satisfied when $d_{int,j}/c_0 > \min(d_{i,j}/c_0)$. In this case, a gap of $S_3$ with $T_{S_3} \geq d_{int,j}/c_0 - \min(d_{i,j}/c_0)$ is added before the TX window, as shown in Fig. 3(c), which satisfies

$$d_{int,j}/c_0 \leq T_{S_1} \leq T_{S_3} + \min(d_{i,j}/c_0). \qquad (13)$$

***Multiplexing using Cyclic Shifts:*** This paper proposes a novel multiplexing scheme to deal with the interference among sensing signals of different TX BSs. Fig 3(d) shows an example of OFDM sensing OFDM symbols with different cyclic shifts. Taking the OFDM symbol with cyclic shift 0 as the base symbol, the symbol with cyclic shift 1~3 is obtained by cyclically shifting the base symbol by $T_{sym}/4$, $T_{sym}/2$, and $3T_{sym}/4$, where $T_{sym}$ is the time length of an OFDM symbol. Then, CP is accordingly added for each symbol. For this multiplexing method, the maximum sensing dynamic range is limited to $T_{sym}/4$. The receiver side detects the delays using the base symbol, and signals from different BSs can be distinguished as they use different delay ranges, e.g., $0.6T_{sym}$ is from cyclic shift 2. Note that the maximum sensing dynamic range has already been limited by the sensing range and CP length. This method reuses the unused cyclic shifts due to coverage and CP limitations for multiplexing. No matter how many sensing signals there are, the RX BSs only require to treat them as one and estimate the parameters of the base symbol. TX BSs are automatically distinguished using the sensing results.

This paper also designs a signaling procedure to realize NCS-MM, which is illustrated in Fig. 4. The sensing function (SF) sends the sensing request, TX resource allocation, and cyclic shift allocation to the BSs in the NCS set. The TX BSs first send an initial sensing signal to the RX BSs, and interference of different types is detected in the sensing RX window and the following uplink slots. Also, TA is measured with this initial sensing signal. With interference and TA measurements, the RX BSs make the decision of whether a reallocation of resources is needed. If needed, the interference and TA information will be sent to the SF, and the SF reallocates the resources according to the information of all RX BSs. This step is only required in NCS initialization and updating, and afterwards, the historical information can be used to set up an appropriate resource allocation. Also, if the GP used in communication is not long enough to satisfy the requirements, SF can configure a larger GP.

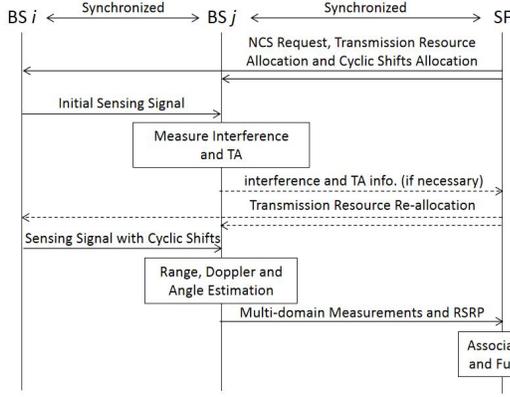

**Fig. 4.** The signaling procedure of the proposed NCS protocol.

After the confirmation of allocated resources in NCS, the TX BSs transmit the sensing signal with different cyclic shifts to sense the targets. The RX BSs will process the receiving signals and estimate the range, Doppler, and 2D AoAs. Apart from the MM estimations, the parameter reflecting the estimation quality should be transmitted, e.g., echo power of the corresponding target. With the estimated parameters and estimation quality, SF can associate and fuse the MM estimations of each TX-RX pair into target positions and velocities.

## III. Performance Limits of NCS

### A. CRLB of MM

The CRLB [41] provides a lower bound for the mean-square-error (MSE) of the unbiased estimator. Given a parameter vector $\mathbf{\Theta} = [\theta_0, \theta_1, \ldots, \theta_{A-1}]^T \in \mathbb{R}^{A \times 1}$, the MSE of unbiased estimation satisfies

$$\text{var}\left([\hat{\mathbf{\Theta}}]_a\right) \geq \text{CRLB}\left([\hat{\mathbf{\Theta}}]_a\right) = [\mathbf{J}^{-1}(\mathbf{\Theta})]_{a,a}, a = 0 \sim A-1, \quad (14)$$

where $\mathbf{J}(\mathbf{\Theta})$ is the FIM.

For real-number parameter estimations of the observation vector $\mathbf{s}(\mathbf{\Theta})$, the FIM $\mathbf{J}(\mathbf{\Theta})$ [41] under the complex-number addictive white Gaussian noise is

$$[\mathbf{J}(\mathbf{\Theta})]_{i',j'} = \frac{2}{\sigma_w^2} \text{Re}\left(\sum_n \left[\frac{\partial [\mathbf{s}(\mathbf{\Theta})]_n}{\partial \theta_{i'}}\right]^* \left[\frac{\partial [\mathbf{s}(\mathbf{\Theta})]_n}{\partial \theta_{j'}}\right]\right), \quad (15)$$

Since the original FIM $\mathbf{J}(\mathbf{\Theta})$ can be difficult to calculate, the chain rule [41] can be applied as

$$\mathbf{J}(\mathbf{\Theta}) = \left(\frac{\partial \mathbf{\Psi}}{\partial \mathbf{\Theta}}\right)^T \mathbf{J}(\mathbf{\Psi}) \frac{\partial \mathbf{\Psi}}{\partial \mathbf{\Theta}}. \quad (16)$$

In related works [33], [34], the CRLB of position and velocity is derived based on the assumption of an known covariance matrix of MM. In the covariance matrix, the azimuth and elevation angles from the BS to the target are used. However, it is complex to directly estimate these two angles, and the error covariance matrix is hard to obtain as the two AoA estimations are coupled. This paper uses decoupling estimations in each dimension using the multi-domain decoupling model in Equation (9).

Assume that $\mathbf{n} = [n_0, n_1, n_2, n_3]^T, n_a = 0,1,\ldots, N_a - 1$. The multi-dimension channel model is written as

$$h_{i,j,k}[n] = A_{i,j,k} e^{j\phi_0^{i,j,k}} \prod_{a=0}^{3} e^{j2\pi f_a^{i,j,k} n_a} + \omega[n], \quad (17)$$

where $\omega[n]$ is the complex-Gaussian channel estimation error with a covariance of $\sigma_\omega^2$.

There are 6 unknown parameters in this multi-dimension model, and let $\mathbf{\Psi}_{i,j,k} = [A_{i,j,k}, \phi_{i,j,k}, f_{i,j,k}^{\text{DM0}}, f_{i,j,k}^{\text{DM1}}, f_{i,j,k}^{\text{DM2}}, f_{i,j,k}^{\text{DM3}}]^T \in \mathbb{R}^{6 \times 1}$. The CRLB of the last four elements determines the MM estimation performance. Substituting the one-dimension sum in Equation (15) with the four-dimension sum, the non-zero elements in FIM $\mathbf{J}(\mathbf{\Psi}_{i,j,k}) \in \mathbb{R}^{6 \times 6}$ are calculated by

$$[\mathbf{J}(\mathbf{\Psi}_{i,j,k})]_{0,0} = \frac{2}{\sigma_w^2} N_0 N_1 N_2 N_3,$$
$$[\mathbf{J}(\mathbf{\Psi}_{i,j,k})]_{1,1} = \frac{2}{\sigma_w^2} A_{i,j,k}^2 N_0 N_1 N_2 N_3,$$
$$[\mathbf{J}(\mathbf{\Psi}_{i,j,k})]_{i'+2,1} = [\mathbf{J}(\mathbf{\Psi})]_{2,i} = \frac{2}{\sigma_w^2} A_{i,j,k}^2 \frac{N_0 N_1 N_2 N_3}{N_{i'}} \sum_{n_{i'}} 2\pi n_{i'},$$
$$[\mathbf{J}(\mathbf{\Psi}_{i,j,k})]_{i'+2,i'+2} = \frac{2}{\sigma_w^2} A_{i,j,k}^2 \frac{N_0 N_1 N_2 N_3}{N_{i'}} \sum_{n_{i'}} (2\pi n_{i'})^2,$$
$$[\mathbf{J}(\mathbf{\Psi}_{i,j,k})]_{i'+2,j'+2} = \frac{2}{\sigma_w^2} A_{i,j,k}^2 \frac{N_0 N_1 N_2 N_3}{N_{i'} N_{j'}} \sum_{n_{i'}} 2\pi n_{i'} \sum_{n_{j'}} 2\pi n_{j'},$$
$$(18)$$

where $i', j' = 0,1,2,3$, and $i' \neq j'$. Other elements in $\mathbf{J}(\mathbf{\Psi})$ are zero due to the Re{·} operation. The matrix inverse of $\mathbf{J}(\mathbf{\Psi})$ is hard to express in the closed form. However, the CRLB of MM can be calculated in a much simpler way.

As the frequencies are decoupled in the multi-domain decoupling model, the CRLB can be calculated in each dimension. For dimension $a$, the power of other dimensions is accumulated via assuming that the target echo energy is coherently added and the noise is non-coherently added. The accumulated single-dimension channel model in dimension $a$ for the $k$-th target is

$$\begin{aligned}
h_{i,j,k}^{\text{DM}a}[n_a] &= G_{\sim a} A_{i,j,k} e^{j\phi_{i,j,k}} e^{j2\pi f_{i,j,k}^{\text{DM}a} n_a} + \sqrt{G_{\sim a}} \omega[n_a] \\
&\triangleq \tilde{A}_{i,j,k}^{\text{DM}a} e^{j\phi_{i,j,k}} e^{j2\pi f_{i,j,k}^{\text{DM}a} n_a} + \sqrt{G_{\sim a}} \omega[n_a],
\end{aligned} \quad (19)$$

where $G_{\sim a} = \prod_{i \neq a} N_i$, $\tilde{A}_{i,j,k}^{\text{DM}a} \triangleq \prod_{i \neq a} N_i A_{i,j,k}$, and $\omega[n_a]$ is the complex-Gaussian channel estimation error with a covariance of $\sigma_\omega^2$.

There are only three unknown parameters to estimate in this model, which is written as $\mathbf{\Psi}_{i,j,k}^{\text{DM}a} = [\tilde{A}_{i,j,k}^{\text{DM}a}, \phi_{i,j,k}, f_{i,j,k}^{\text{DM}a}]^T$. The decoupling FIM is

$$\mathbf{J}(\mathbf{\Psi}_{i,j,k}^{\text{DM}a}) = \frac{2}{G_{\sim a}\sigma_\omega^2} \begin{bmatrix} N_a & 0 & 0 \\ 0 & (G_{\sim a}A_{i,j,k})^2 N_a & (G_{\sim a}A_{i,j,k})^2 \Sigma_{n_a} 2\pi n_a \\ 0 & (G_{\sim a}A_{i,j,k})^2 \Sigma_{n_a} 2\pi n_a & (G_{\sim a}A_{i,j,k})^2 \Sigma_{n_a} (2\pi n_a)^2 \end{bmatrix}$$

$$= \frac{2}{G_{\sim a}\sigma_\omega^2} \begin{bmatrix} N_a & 0 & 0 \\ 0 & (G_{\sim a}A_{i,j,k})^2 N_a & \pi(G_{\sim a}A_{i,j,k})^2 N_a(N_a-1) \\ 0 & \pi(G_{\sim a}A_{i,j,k})^2 N_a(N_a-1) & 4\pi^2 (G_{\sim a}A_{i,j,k})^2 \frac{N_a(N_a-1)(2N_a-1)}{6} \end{bmatrix}. \quad (20)$$

The matrix inverse of $\mathbf{J}(\mathbf{\Psi}_{i,j,k})$ can be written in closed form. As the frequency is what we care about in wireless sensing, the related element in $\mathbf{J}^{-1}(\mathbf{\Psi}_{i,j,k}^{\text{DM}a})$, or the CRLB, of the decoupled frequency in dimension $a$ is

$$\text{CRLB}(\hat{f}_{i,j,k}^{\text{DM}a}) = [\mathbf{J}^{-1}(\mathbf{\Psi}_{i,j,k}^{\text{DM}a})]_{2,2} = \frac{3\sigma_\omega^2}{2\pi^2 A_{i,j,k}^2 N_0 N_1 N_2 N_3 (N_a^2 - 1)}. \quad (21)$$

Now, a unified CRLB expression for different multi-domain frequencies is obtained. The CRLB of MM in the TX-RX pair $(i,j)$ can be easily obtained using the relationship in Equation (10), which multiplies $\text{CRLB}(\hat{f}_{i,j,k}^{\text{DM}a})$, $a = 0,1,2,3$ with $c_0^2/\Delta f^2$, $c_0^2/(f_c T)^2$, 4 and 4, respectively.

Although the accumulated single-dimension model is much simpler, the concern is that there is an underlying assumption of a perfect target echo energy accumulation. This paper further proves the correctness of the simplified model. Denote the right-bottom 4×4 matrix of $\mathbf{J}^{-1}(\mathbf{\Psi}_{i,j,k})$ as $[\mathbf{J}^{-1}(\mathbf{\Psi}_{i,j,k})]_{4\times 4}$. This paper has the following proposition:

*Proposition 1:* The CRLB of MM derived from the complete FIM matrix $\mathbf{J}(\mathbf{\Psi}_{i,j,k})$ equals that from the decoupling FIM $\mathbf{J}(\mathbf{\Psi}_{i,j,k}^{\text{DM}a})$, which can be stated by

$$[\mathbf{J}^{-1}(\mathbf{\Psi}_{i,j,k})]_{4\times 4} = \text{diag}(J_{i,j,k}^{\text{DM0}}, J_{i,j,k}^{\text{DM1}}, J_{i,j,k}^{\text{DM2}}, J_{i,j,k}^{\text{DM3}}), \quad (22)$$

where $J_{i,j,k}^{\text{DM}a} = [\mathbf{J}^{-1}(\mathbf{\Psi}_{i,j,k}^{\text{DM}a})]_{2,2}$.

The proof can be found in Appendix A. This proof shows that the CRLB result of multi-domain frequencies can be calculated by the simpler accumulated single-dimension model with the help of the multi-domain decoupling model.

The CRLB of the frequency in each dimension can be compared using the united form in Equation (21). In practice, the sub-carrier number $N$ can be up to several thousand. However, the antenna number in each dimension, $L_x$ or $L_y$, can only be a couple of tens, even when assuming a thousand antennas in a panel. That will lead to a MSE mismatch between range and cross-range. This problem also shows the necessity of NCS, as the range estimations of multiple views can be fused to reduce the uncertainty in the cross range.

### B. CRLB of Positions and Velocities in NCS

Assume $\mathbf{\Omega}_k = [\mathbf{\Omega}_k^{\text{DM0}^T}, \mathbf{\Omega}_k^{\text{DM1}^T}, \mathbf{\Omega}_k^{\text{DM2}^T}, \mathbf{\Omega}_k^{\text{DM3}^T}]^T \in \mathbb{R}^{4IJ\times 1}$ with $\mathbf{\Omega}_k^{\text{DM}a} = [f_{0,JS,k}^{\text{DM}a}, \ldots, f_{I-1,J+JS-1,k}^{\text{DM}a}]^T \in \mathbb{R}^{IJ\times 1}$. This paper uses $\mathbf{\Omega}_k$ to estimate the 3D position and the velocity of the $k$-th target, $\mathbf{\Theta}_k = [\mathbf{t}_k, \dot{\mathbf{t}}_k]^T$. The FIM of the decoupled frequencies can be recovered via the CRLB of $[\mathbf{\Psi}_{i,j,k}]_{2\sim 5}$ as

$$\mathbf{J}([\mathbf{\Psi}_{i,j,k}]_{2\sim 5}) = ([\mathbf{J}^{-1}(\mathbf{\Psi}_{i,j,k})]_{4\times 4})^{-1}, \quad (23)$$

Note that the result is diagonal according to Equation (19). Then, we can have

$$\mathbf{J}(\mathbf{\Omega}_k) = \text{diag}(\mathbf{J}(\mathbf{\Omega}_k^{\text{DM0}}), \mathbf{J}(\mathbf{\Omega}_k^{\text{DM1}}), \mathbf{J}(\mathbf{\Omega}_k^{\text{DM2}}), \mathbf{J}(\mathbf{\Omega}_k^{\text{DM3}})), \quad (24)$$

where $\mathbf{J}(\mathbf{\Omega}_k^{\text{DM}a}) = \text{diag}(J_{0,JS,k}^{\text{DM}a}, \ldots, J_{I-1,J+JS-1,k}^{\text{DM}a})$ as the measurements of different transmitting pairs are independent.

Using Equations (14) and (16), the CRLB of $\mathbf{\Theta}_k$ is

$$\text{CRLB}(\mathbf{\Theta}_k) = \mathbf{J}^{-1}(\mathbf{\Theta}_k) = \left(\frac{\partial \mathbf{\Omega}_k}{\partial \mathbf{\Theta}_k} \mathbf{J}(\mathbf{\Omega}_k) \left(\frac{\partial \mathbf{\Omega}_k}{\partial \mathbf{\Theta}_k}\right)^T\right)^{-1}, \quad (25)$$

The Jacobian matrix $\partial \mathbf{\Psi}_k / \partial \mathbf{\Theta}_k$ is calculated by

$$\frac{\partial \mathbf{\Omega}_k}{\partial \mathbf{\Theta}_k} = \left[\left(\frac{\partial \mathbf{\Omega}_k^{\text{DM0}}}{\partial \mathbf{\Theta}_k}\right)^T, \left(\frac{\partial \mathbf{\Omega}_k^{\text{DM1}}}{\partial \mathbf{\Theta}_k}\right)^T, \left(\frac{\partial \mathbf{\Omega}_k^{\text{DM2}}}{\partial \mathbf{\Theta}_k}\right)^T, \left(\frac{\partial \mathbf{\Omega}_k^{\text{DM3}}}{\partial \mathbf{\Theta}_k}\right)^T\right]^T,$$

$$\frac{\partial \mathbf{\Omega}_k^{\text{DM}a}}{\partial \mathbf{\Theta}_k} = \left[\frac{\partial f_{0,JS,k}^{\text{DM}a}}{\partial \mathbf{\Theta}_k}, \ldots, \frac{\partial f_{I-1,J+JS-1,k}^{\text{DM}a}}{\partial \mathbf{\Theta}_k}\right]^T,$$

$$\frac{\partial f_{0,JS,k}^{\text{DM0}}}{\partial \mathbf{\Theta}_k} = \left[-\frac{\Delta f}{c_0} \cdot (\boldsymbol{\rho}_{\mathbf{t}_k,\mathbf{b}_i} + \boldsymbol{\rho}_{\mathbf{t}_k,\mathbf{b}_j})^T, \mathbf{0}_{1\times 3}\right],$$

$$\frac{\partial f_{0,JS,k}^{\text{DM1}}}{\partial \mathbf{\Theta}_k} = \left[\frac{f_c T}{c_0} \cdot (\boldsymbol{\xi}_{\mathbf{t}_k,\mathbf{b}_i,\mathbf{t}} + \boldsymbol{\xi}_{\mathbf{t}_k,\mathbf{b}_j,\mathbf{t}})^T, \frac{f_c T}{c_0} \cdot (\boldsymbol{\rho}_{\mathbf{t}_k,\mathbf{b}_i} + \boldsymbol{\rho}_{\mathbf{t}_k,\mathbf{b}_j})^T\right],$$

$$\frac{\partial f_{0,JS,k}^{\text{DM2}}}{\partial \mathbf{\Theta}_k} = \left[\frac{1}{2}\boldsymbol{\xi}_{\mathbf{t}_k,\mathbf{b}_j,\mathbf{x}_i}^T, \mathbf{0}_{1\times 3}\right], \frac{\partial f_{0,JS,k}^{\text{DM3}}}{\partial \mathbf{\Theta}_k} = \left[\frac{1}{2}\boldsymbol{\xi}_{\mathbf{t}_k,\mathbf{b}_j,\mathbf{x}_j}^T, \mathbf{0}_{1\times 3}\right], \quad (26)$$

where $\boldsymbol{\xi}_{\mathbf{a},\mathbf{b},\mathbf{c}} \triangleq (\mathbf{I}_3 - \boldsymbol{\rho}_{\mathbf{a},\mathbf{b}} \boldsymbol{\rho}_{\mathbf{a},\mathbf{b}}^T) \mathbf{c} / \|\mathbf{a} - \mathbf{b}\|$.

Compared to the proposed CRLB of Equations (25) and (26), existing works [33],[34] used the following CRLB

$$\text{CRLB}(\mathbf{\Theta}_k) = \left(\frac{\partial \mathbf{\Omega}_k}{\partial \mathbf{\Theta}_k} \mathbf{Q}_m^{-1} \left(\frac{\partial \mathbf{\Omega}_k}{\partial \mathbf{\Theta}_k}\right)^T\right)^{-1}, \quad (27)$$

where $\mathbf{Q}_m$ is assumed to be a known diagonal covariance matrix. The difference is that the existing work directly uses $\mathbf{J}(\mathbf{\Omega}_k) = \mathbf{Q}_m^{-1}$. However, $\mathbf{Q}_m$ is not diagonal due to the coupling of azimuth angle and elevation angle [33],[34]. That is why this paper proposes to use decoupling 2D AoA.

In some NCS cases, the 3D velocity cannot be recovered. Similar to Equation (25), the position-only CRLB is

$$\text{CRLB}(\mathbf{t}_k) = \mathbf{J}^{-1}(\mathbf{t}_k) = \left(\frac{\partial \mathbf{\Omega'}_k}{\partial \mathbf{t}_k} J(\mathbf{\Omega'}_k) \left(\frac{\partial \mathbf{\Omega'}_k}{\partial \mathbf{t}_k}\right)^T\right)^{-1}, \quad (28)$$

where $\mathbf{\Omega'}_k = [\mathbf{\Omega}_k^{\text{DM0}^T}, \mathbf{\Omega}_k^{\text{DM2}^T}, \mathbf{\Omega}_k^{\text{DM3}^T}]^T$. $\mathbf{J}(\mathbf{\Omega'}_k)$ and $\frac{\partial \mathbf{\Omega'}_k}{\partial \mathbf{t}_k}$ can be obtained using relating parts in Equations (24) and (26).

## IV. ESTIMATION AND FUSION ALGORITHMS

The processing flow of NCS-MM includes three major steps: the MM estimation for each TX-RX pair, the target association across different pairs, and finally the MM fusion for each target. This paper focuses on the first and last ones. For the target association, the targets observed by different TX-RX pairs can be seen as one target if their distances are short.

### A. Grid-free MM Estimation

With the proposed decoupling model, the estimation problem is to obtain the multi-domain frequency estimations of Equation (9). As the algorithm is general for each $(i,j)$ pair, the index $(i,j)$ is omitted in this sub-section, and the multiple targets' channel information with noise is modeled as

$$\mathbf{y} = \sum_{i=0}^{k-1} A_{i,j,k} e^{j\phi_{i,j,k}} \mathbf{\Lambda}_{i,j,k}^{DM3} \otimes \mathbf{\Lambda}_{i,j,k}^{DM2} \otimes \mathbf{\Lambda}_{i,j,k}^{DM1} \otimes \mathbf{\Lambda}_{i,j,k}^{DM0} + \mathbf{n}$$

$$\triangleq \sum_{i=0}^{k-1} g_k \mathbf{a}_0(f_{k,0}) \otimes \mathbf{a}_1(f_{k,1}) \otimes \ldots \otimes \mathbf{a}_{A-1}(f_{k,A-1}) + \mathbf{n}$$

$$\triangleq \sum_{i=0}^{k-1} g_k \mathbf{a}_{AD}(\mathbf{f}_k) + \mathbf{n}, \tag{29}$$

where $\mathbf{f}_k = [f_{k,0}, f_{k,1}, \ldots, f_{k,A-1}]^T \in \mathbb{R}^{A\times 1}$, and $\mathbf{n}$ is the additive white Gaussian noise.

This paper considers grid-less methods to estimate multi-domain frequencies. The different dimensions are decoupled in Equation (29), and each dimension can be processed using the same method. Therefore, this paper proposes arbitrary-dimension NOMP (AD-NOMP) to deal with the decoupled arbitrary-dimension frequency estimations. A uniform planar array (UPA)-NOMP [42] has been proposed to estimate range and 2D AoAs, and the 2D AoAs in UPA-NOMP are coupled. The advantage of the proposed AD-NOMP is that is can deal with any combination of range, Doppler, and 2D AoAs as multi-domain decoupling model is used.

The first step of AD-NOMP is to find an initial MM estimation close to the true values of MM. However, it is very complex to implement conventional grid-search method [36] in high dimensions. This paper proposes an efficient dimension-wise search method. For dimension $a$, the estimation of detected dimension 0~$a$-1 is coherently added, and the estimation of undetected dimensions is non-coherently added using

$$\max_{f_{k,a} = \hat{f}_{k,a}} \mathbf{C}_2 \left| \mathbf{C}_1 \left( \mathbf{a}_k \left( [\hat{f}_{k,0}, \ldots \hat{f}_{k,a-1}, f_{k,a}, 0, \ldots, 0]^T \right) \odot \mathbf{y} \right) \right|^2, \tag{30}$$

where $\mathbf{C}_1 = \mathbf{1}_{N_a N_{bf}} \otimes \mathbf{I}_{N_{af}} \in \mathbb{R}^{N_{af} \times N_{bf} N_{af} N_a}$, $N_{bf} = \prod_{i=0}^{a-1} N_i$, $N_{af} = \prod_{i=a+1}^{A-1} N_i$, and $\mathbf{C}_2 = \text{blkdiag}\{\mathbf{1}_{N_2}^T, \ldots, \mathbf{1}_{N_2}^T\} \in \mathbb{R}^{1\times N_{af}}$.

After the initial estimation, the refinement is done using Newton iterations. The target function is

$$f(\mathbf{f}_k) = \left| \mathbf{a}_k^H(\mathbf{f}_k) \mathbf{y} \right|^2 = \mathbf{a}_k^H(\mathbf{f}_k) \mathbf{y} \mathbf{y}^H \mathbf{a}_k(\mathbf{f}_k), \tag{31}$$

The Newton iterations are done using

$$\hat{\mathbf{f}}_k = \hat{\mathbf{f}}_k - \dot{f}(\mathbf{f}_k)/\mathbf{H}_f(\mathbf{f}_k), \tag{32}$$

where

$$\dot{f}(\mathbf{f}_k) = 2\mathcal{R}\left( \left[ \mathbf{a}^H(\mathbf{f}_k) \mathbf{y} \mathbf{y}^H \frac{\partial \mathbf{a}(\mathbf{f}_k)}{\partial [f_k]_0}, \ldots, \mathbf{a}^H(\mathbf{f}_k) \mathbf{y} \mathbf{y}^H \frac{\partial \mathbf{a}(\mathbf{f}_k)}{\partial [f_k]_{A-1}} \right]^T \right), \tag{33}$$

$$\mathbf{H}_f(\mathbf{f}_k) = \begin{bmatrix} \frac{\partial f(\mathbf{f}_k)}{\partial [f_k]_0^2} & \frac{\partial f(\mathbf{f}_k)}{\partial [f_k]_0 \partial [f_k]_1} & \cdots & \frac{\partial f(\mathbf{f}_k)}{\partial [f_k]_0 \partial [f_k]_{A-1}} \\ \frac{\partial f(\mathbf{f}_k)}{\partial [f_k]_1 \partial [f_k]_0} & \frac{\partial f(\mathbf{f}_k)}{\partial [f_k]_1^2} & \cdots & \frac{\partial f(\mathbf{f}_k)}{\partial [f_k]_1 \partial [f_k]_{A-1}} \\ \vdots & \vdots & \ddots & \vdots \\ \frac{\partial f(\mathbf{f}_k)}{\partial [f_k]_{A-1} \partial [f_k]_0} & \frac{\partial f(\mathbf{f}_k)}{\partial [f_k]_{A-1} \partial [f_k]_1} & \cdots & \frac{\partial f(\mathbf{f}_k)}{[f_k]_{A-1}^2} \end{bmatrix},$$

$$\frac{\partial f(\mathbf{f}_k)}{\partial [f_k]_a \partial [f_k]_b} = 2\mathcal{R}\left( \mathbf{a}^H(\mathbf{f}_k) \mathbf{y} \mathbf{y}^H \frac{\partial \mathbf{a}(\mathbf{f}_k)}{\partial [f_k]_a \partial [f_k]_b} + \left( \frac{\partial \mathbf{a}(\mathbf{f}_k)}{\partial [f_k]_b} \right)^H \mathbf{y} \mathbf{y}^H \frac{\partial \mathbf{a}(\mathbf{f}_k)}{\partial [f_k]_a} \right), \tag{34}$$

are the derivative vector and Hessian matrix of $f(\mathbf{f}_k)$. As the Hessian matrix can be rank-deficient, our algorithm can keep the diagonal elements. The physical meaning of keeping the diagonal elements is to independently refine each dimension, which also works for the decoupling frequencies.

Apart from the arbitrary dimension, another difference from NOMP [36] is the estimation of $g_k$ does not require to be calculated in Newton iterations. The remaining steps are similar to NOMP, and the procedure is in Algorithm 1. Note that in iterations, $\mathbf{y}$ in (30) is replaced by the residual of

$$\mathbf{y}_r(\mathcal{P}_{m-1}) = \mathbf{y} - \sum_{l=0}^{m-1} \hat{g'}_l \mathbf{a}(\hat{\mathbf{f'}}_l), \tag{35}$$

where $\mathcal{P}_{m-1} = \{(\hat{g'}_l, \hat{\mathbf{f'}}_l)\}, l = 0, 1, \ldots, m-1$.

| Algorithm 1: AD-NOMP |
|---|
| Input: |
| Output: |
| 0: m = 0, $\mathcal{P}_0 = \{\}$ |
| 1: while $|\mathbf{y}_r(\mathcal{P}_{m-1})|^2 > P_{th}$ or $m \leq$ estimate user number |
| 2:   m = m+1 |
| 3:   Find a new $\hat{\mathbf{f}}$ using Equation (30) and $\mathbf{y}_r(\mathcal{P}_{m-1})$ |
| 4:   Refine $\hat{\mathbf{f}}$ to obtain $\hat{\mathbf{f}}'$ using Equation (31) |
| 5:   Calculate $\hat{g}'$ using $\hat{g}' = a$ |
| 5:   $\mathcal{P}'_m = \mathcal{P}_m \cup \{(\hat{g}', \hat{\mathbf{f}}')\}$ |
| 6:    For each $\{(\hat{g}, \hat{\mathbf{f}})\} \in \mathcal{P}'_m$, using Equation (30) and $\mathbf{y}_r(\mathcal{P}'_m \setminus (\hat{g}, \hat{\mathbf{f}}))$ to refine and obtain a updated $\mathcal{P}''_m$ |
| 7:   Update $\hat{g}$ in $\mathcal{P}''_m$ by LS using $[\mathbf{a}(\mathbf{f}_0), \ldots, \mathbf{a}(\mathbf{f}_{m-1})]^\dagger \mathbf{y}$ and obtain a new $\mathcal{P}_m$ |
| 8: end |

### B. Closed-Form MM Fusion in FD-NCS

The MM estimations from AD-NOMP require to be associated with different targets. As each TX-RX pair can generate the estimated position, the targets of the closest estimated positions in different pairs are associated to one target. After association, the measurements of all pairs $\hat{\mathbf{m}}_k = [\hat{\mathbf{r}}_k^T, \hat{\mathbf{r}}_k^T, \hat{\mathbf{c}}_{\alpha_k}^T, \hat{\mathbf{c}}_{\beta_k}^T]^T \in \mathbb{R}^{4IJ \times 1}$ are used for NCS fusion, where $\hat{\mathbf{c}}_\alpha$ denotes the estimations of $\cos\alpha$.

This paper first considers the DoF of MM in the FD-NCS scenario. In FD-NCS, a lot of TX-RX pairs can be obtained, which is several times of the number of BSs as $L = IJ = IN_{BS}$. As $L$ also decides the number of equations, a large $L$ leads to large-scale equations which has a high computational complexity. To solve this problem, this paper proposes to reduce the number of equations and only keep the necessary equations according to the DoF. A TX-RX pair has two trips, and the DoF of independent trips is no more than $N_{BS}$, as the set $[d_{0,k}, d_{1,k}, \ldots, d_{N_{BS}-1,k}]$ can generate all TX-RX distances, and so can the Doppler frequencies. In addition, the 2D angle estimation is in the view of the receiving BS, which means there are $J$ independent estimations instead of $IJ$ for each angle.

The relationship between the independent parameters $\boldsymbol{\mu}_k = [\boldsymbol{d}_k^T, \dot{\boldsymbol{d}}_k^T, \boldsymbol{c}_{\alpha_k}^T, \boldsymbol{c}_{\beta_k}^T]^T \in \mathbb{R}^{(2N_{BS}+2J)\times 1}$ and the measurements $\widehat{\boldsymbol{m}}_k$ is modeled as

$$-\Delta\boldsymbol{m}_k = \widehat{\boldsymbol{m}}_k - \mathbf{T}_1\boldsymbol{\mu}_k, \tag{36}$$

where $\mathbf{T}_1 \triangleq blkdiag(\mathbf{T}_A, \mathbf{T}_A, \mathbf{T}_B, \mathbf{T}_B) \in \mathbb{R}^{4IJ\times(2N_{BS}+2J)}$. The $l$-th row of $\mathbf{T}_A \in \mathbb{R}^{N_{BS}\times IJ}$ and $\mathbf{T}_B \in \mathbb{R}^{J\times IJ}$ is obtained from the summing relationship in the $l$-th $(i,j)$ pair. The non-zero elements in $\mathbf{T}_A$ and $\mathbf{T}_B$ are calculated by

$$\begin{aligned}
[\mathbf{T}_A]_{l,i} &= [\mathbf{T}_A]_{l,j} = 1, \mathrm{m}l = iI + j - J_S \text{ and } i \neq j, \\
[\mathbf{T}_A]_{l,i} &= 2, l = iI + j - J_S \text{ and } i = j, \\
[\mathbf{T}_B]_{l,j} &= 1, l = iI + j - J_S.
\end{aligned} \tag{37}$$

The minimum variance unbiased estimator [41] for Equation (36) is to use a WLS of

$$\widehat{\boldsymbol{\mu}}_k = \left(\mathbf{T}_1^H \mathbf{W}_{m_k} \mathbf{T}_1\right)^{-1} \mathbf{T}_1^H \mathbf{W}_{m_k} \widehat{\boldsymbol{m}}_k. \tag{38}$$

where

$$\begin{aligned}
\mathbf{W}_{m_k} &= \mathbf{Q}_{m_k}^{-1} = E[\Delta\boldsymbol{m}_k \Delta\boldsymbol{m}_k^T]^{-1} \\
&= diag\left(\frac{c_0^2}{\Delta f^2}\mathbf{J}(\Omega_k^{DM0}), \frac{c_0^2}{(f_cT)^2}\mathbf{J}(\Omega_k^{DM1}), 4\mathbf{J}(\Omega_k^{DM2}), 4\mathbf{J}(\Omega_k^{DM3})\right).
\end{aligned} \tag{39}$$

The error covariance matrix of $\widehat{\boldsymbol{\mu}}_k$ is

$$E[\Delta\boldsymbol{\mu}_k \Delta\boldsymbol{\mu}_k^T] = \left(\mathbf{T}_1^H \mathbf{W}_{m_k} \mathbf{T}_1\right)^{-1}, \tag{40}$$

With Equations (36)~(40), the redundancy of measurements is removed, and the number of measurements is reduced from $4IN_{BS}$ to $(2N_{BS}+2J)$. Also, as DoF is not reduced, performance can still be ensured.

For the distance measurements, we can have

$$d_{i_0,k}^2 = (\mathbf{b}_{i_0} - \mathbf{t}_k)^T(\mathbf{b}_{i_0} - \mathbf{t}_k) = \mathbf{b}_{i_0}^T\mathbf{b}_{i_0} - 2\mathbf{b}_{i_0}^T\mathbf{t}_k + \mathbf{t}_k^T\mathbf{t}_k, \tag{41}$$

which can be transformed into

$$2\mathbf{b}_{i_0}^T\mathbf{t}_k = \mathbf{b}_{i_0}^T\mathbf{b}_{i_0} - d_{i_0,k}^2 + \mathbf{t}_k^T\mathbf{t}_k, \tag{42}$$

Take BS 0 as the reference station, subtract Equation (42) when $i = 1, ..., N_{BS}-1$ by that of $i = 0$, and

$$2(\mathbf{b}_{i_1} - \mathbf{b}_0)^T\mathbf{t}_k = \mathbf{b}_{i_1}^T\mathbf{b}_{i_1} - \mathbf{b}_0^T\mathbf{b}_0 - d_{i_1,k}^2 + d_{0,k}^2, \tag{43}$$

For the velocity estimations, the linear equations can be formatted using the derivative of Equation (43) as

$$2(\mathbf{b}_{i_1} - \mathbf{b}_0)^T\dot{\mathbf{t}}_k = -2d_{i_1,k}\dot{d}_{i_1,k} + 2d_{0,k}\dot{d}_{0,k}, \tag{44}$$

For the angle estimations, the following linear equations is obtained using some simple manipulations:

$$\begin{aligned}
\mathbf{x}_j^T\mathbf{t}_k &= \mathbf{x}_j^T\mathbf{b}_j + d_{j,k}c_{\alpha_{j,k}}, \\
\mathbf{y}_j^T\mathbf{t}_k &= \mathbf{y}_j^T\mathbf{b}_j + d_{j,k}c_{\beta_{j,k}}.
\end{aligned} \tag{45}$$

Stacking Equations (43), (44) and (45) for $i_1 = 1, \ldots, N_{BS}-1$ and $j = J_S, \ldots, J_S + J - 1$ into large-scale linear equations and representing the true measurements with the estimated ones, e.g $d_{k,i_1} = \widehat{d}_{k,i_1} + \Delta d_{k,i_1}$,

$$\boldsymbol{\varepsilon}_1 = \mathbf{T}_2 \Delta\boldsymbol{\mu}_k = \widehat{\boldsymbol{b}}_1 - \mathbf{A}_1\boldsymbol{\Theta}_1, \tag{46}$$

where $\widehat{\boldsymbol{b}}_1$ is obtained using

$$\begin{aligned}
[\widehat{\boldsymbol{b}}_1]_{i_1-1} &= \mathbf{b}_{i_1}^T\mathbf{b}_{i_1} - \mathbf{b}_0^T\mathbf{b}_0 - \widehat{d}_{k,i_1}^2 + \widehat{d}_{k,0}^2, \\
[\widehat{\boldsymbol{b}}_1]_{N_{BS}+i_1-2} &= -2\widehat{d}_{k,i_1}\widehat{\dot{d}}_{k,i_1} + 2\widehat{d}_{k,0}\widehat{\dot{d}}_{k,0}, \\
[\widehat{\boldsymbol{b}}_1]_{2N_{BS}+j-J_S-2} &= \mathbf{x}_j^T\mathbf{b}_j + \widehat{d}_{j,k}\widehat{c}_{\alpha_{j,k}}, \\
[\widehat{\boldsymbol{b}}_1]_{3N_{BS}+j-2J_S-2} &= \mathbf{y}_j^T\mathbf{b}_j + \widehat{d}_{j,k}\widehat{c}_{\beta_{j,k}},
\end{aligned} \tag{47}$$

and $\mathbf{A}_1$ is calculated via

$$\begin{aligned}
[\mathbf{A}_1]_{i_1-1,1\sim 6} &= [2(\mathbf{b}_i - \mathbf{b}_0)^T, \mathbf{0}_{1\times 3}], \\
[\mathbf{A}_1]_{N_{BS}+i_1-2,1\sim 6} &= [\mathbf{0}_{1\times 3}, 2(\mathbf{b}_i - \mathbf{b}_0)^T], \\
[\mathbf{A}_1]_{2N_{BS}+j-J_S-2,1\sim 6} &= [\mathbf{x}_j^T, \mathbf{0}_{1\times 3}], \\
[\mathbf{A}_1]_{3N_{BS}+j-2J_S-2,1\sim 6} &= [\mathbf{y}_j^T, \mathbf{0}_{1\times 3}].
\end{aligned} \tag{48}$$

In Equation (46), $\mathbf{T}_2$ is very important to implement WLS, and it can be divided into 4 parts as $\mathbf{T}_2 \triangleq [\mathbf{T}_C^T, \mathbf{T}_D^T, \mathbf{T}_E^T, \mathbf{T}_F^T]^T \in \mathbb{R}^{(2N_{BS}+2J-2)\times(2N_{BS}+2J)}$, where $\mathbf{T}_C, \mathbf{T}_D \in \mathbb{R}^{(N_{BS}-1)\times(2N_{BS}+2J)}$, and $\mathbf{T}_E, \mathbf{T}_F \in \mathbb{R}^{J\times(2N_{BS}+2J)}$. Omitting the small second-order errors, the non-zero elements in these matrices are calculated by

$$\begin{aligned}
[\mathbf{T}_C]_{i_1-1,0} &= 2\widehat{d}_{0,k}, \quad [\mathbf{T}_C]_{i_1-1,i_1} = -2\widehat{d}_{i_1,k}, \\
[\mathbf{T}_D]_{i_1-1,0} &= 2\widehat{\dot{d}}_{0,k}, \quad [\mathbf{T}_D]_{i_1-1,i_1} = -2\widehat{\dot{d}}_{i_1,k}, \\
[\mathbf{T}_D]_{i_1-1,N_{BS}-1} &= 2\widehat{d}_{0,k}, \quad [\mathbf{T}_D]_{i_1-1,N_{BS}-1+i_1} = -2\widehat{d}_{i_1,k}, \\
[\mathbf{T}_E]_{j-J_S,j} &= \widehat{c}_{\alpha_{j,k}}, \quad [\mathbf{T}_E]_{j-J_S,2N_B+j-J_S} = \widehat{d}_{j,k}, \\
[\mathbf{T}_F]_{j-J_S,j} &= \widehat{c}_{\beta_{j,k}}, \quad [\mathbf{T}_F]_{j-J_S,2N_B+J+j-J_S} = \widehat{d}_{j,k},
\end{aligned} \tag{49}$$

The first-stage WLS is

$$\widehat{\boldsymbol{\Theta}}_1 = \left(\mathbf{A}_1^H \mathbf{W}_1 \mathbf{A}_1\right)^{-1} \mathbf{A}_1^H \mathbf{W}_1 \widehat{\boldsymbol{b}}_1, \tag{50}$$

where

$$\mathbf{W}_1 = \mathbf{Q}_{\varepsilon_1}^{-1} = \left(\mathbf{T}_2^H \mathbf{Q}_{\mu_k} \mathbf{T}_2\right)^{-1} = \left(\mathbf{T}_2^H (\mathbf{T}_1^H \mathbf{Q}_{m_k}^{-1} \mathbf{T}_1)^{-1} \mathbf{T}_2\right)^{-1}, \tag{51}$$

The covariance matrix of $\widehat{\boldsymbol{\Theta}}_1$ is

$$\mathbf{Q}_{\Theta_1} = \left(\mathbf{A}_1^H \mathbf{W}_1 \mathbf{A}_1\right)^{-1}, \tag{52}$$

Although WLS gives the best linear unbiased estimations [41], one DoF is lost to construct the linear equations. To further refine the results, a second-stage WLS is used.

For the distance estimations, replacing the true values with the estimations and errors in Equation (42), and omitting the second-order errors,

$$2\mathbf{b}_{i_0}^T(\widehat{\mathbf{t}}_k + \Delta\mathbf{t}_k) \approx \mathbf{b}_{i_0}^T\mathbf{b}_{i_0} - \widehat{d}_{i_0,k}^2 - 2\widehat{d}_{i_0,k}\Delta d_{i_0,k} + \widehat{\mathbf{t}}_k^T\widehat{\mathbf{t}}_k + 2\widehat{\mathbf{t}}_k\Delta\mathbf{t}_k. \tag{53}$$

By some algebraic manipulations, linear equations of position estimation errors can be obtained as

$$2(\mathbf{b}_{i_0} - \widehat{\mathbf{t}}_k)^T \Delta\mathbf{t}_k \approx \mathbf{b}_{i_0}^T\mathbf{b}_{i_0} - \widehat{d}_{i_0,k}^2 + \widehat{\mathbf{t}}_k^T\widehat{\mathbf{t}}_k - 2\mathbf{b}_{i_0}^T\widehat{\mathbf{t}}_k - 2\widehat{d}_{i_0,k}\Delta d_{i_0,k}. \tag{54}$$

For the velocity estimations, the derivative of Equation (42) can be written as

$$2\mathbf{b}_{i_0}^T(\hat{\mathbf{t}}_k + \Delta\mathbf{t}_k) \approx -2\hat{d}_{i_0,k}\hat{d}_{i_0,k} - 2\hat{d}_{i_0,k}\Delta d_{i_0,k} - 2\hat{d}_{i_0,k}\Delta d_{i_0,k} + 2\hat{\mathbf{t}}_k^T\hat{\mathbf{t}}_k + 2\hat{\mathbf{t}}_k^T\Delta\mathbf{t}_k + 2\hat{\mathbf{t}}_k^T\Delta\mathbf{t}_k. \quad (55)$$

Linear equations of position and velocity estimation errors are

$$-2\hat{\mathbf{t}}_k^T\Delta\mathbf{t}_k + 2(2\mathbf{b}_{i_0} - 2\hat{\mathbf{t}}_k)^T\Delta\mathbf{t}_k \approx -2\hat{d}_{i_0,k}\hat{d}_{i_0,k} + 2\hat{\mathbf{t}}_k^T\Delta\mathbf{t}_k - 2\mathbf{b}_{i_0}^T\hat{\mathbf{t}}_k - 2\hat{d}_{i_0,k}\Delta d_{i_0,k} - 2\hat{d}_{i_0,k}\Delta d_{i_0,k}. \quad (56)$$

Based on the angle estimations in Equation (45),

$$\mathbf{x}_j^T\Delta\mathbf{t}_k = \mathbf{x}_j^T\mathbf{b}_j - \mathbf{x}_j^T\hat{\mathbf{t}}_k + \hat{d}_{j,k}\hat{c}_{\alpha_{j,k}} + \hat{c}_{\alpha_{j,k}}\Delta d_{j,k} + \hat{d}_{j,k}\Delta c_{\alpha_{j,k}},$$
$$\mathbf{y}_j^T\Delta\mathbf{t}_k = \mathbf{y}_j^T\mathbf{b}_j - \mathbf{y}_j^T\hat{\mathbf{t}}_k + \hat{d}_{j,k}\hat{c}_{\beta_{j,k}} + \hat{c}_{\beta_{j,k}}\Delta d_{j,k} + \hat{d}_{j,k}\Delta c_{\beta_{j,k}}, \quad (57)$$

Stacking Equations (54), (56) and (57) for $i_0 = 0,\ldots,N_{BS}-1$ and $j = J_S,\ldots,J_S + J - 1$, the second-stage model can be obtained as

$$\boldsymbol{\varepsilon}_2 = \mathbf{T}_3\Delta\boldsymbol{\mu}_k = \hat{\mathbf{b}}_2 - \hat{\mathbf{A}}_2\boldsymbol{\Theta}_2, \quad (58)$$

where $\hat{\mathbf{b}}_2$ is obtained via

$$[\hat{\mathbf{b}}_2]_{i_0} = \mathbf{b}_{i_0}^T\mathbf{b}_{i_0} - \hat{d}_{i_0,k}^2 + \hat{\mathbf{t}}_k^T\hat{\mathbf{t}}_k - 2\mathbf{b}_{i_0}^T\hat{\mathbf{t}}_k,$$
$$[\hat{\mathbf{b}}_2]_{N_{BS}+i_0} = -2\hat{d}_{i_0,k}\hat{d}_{i_0,k} + 2\hat{\mathbf{t}}_k^T\hat{\mathbf{t}}_k - 2\mathbf{b}_{i_0}^T\hat{\mathbf{t}}_k,$$
$$[\hat{\mathbf{b}}_2]_{2N_{BS}+j-J_S} = \mathbf{x}_j^T\mathbf{b}_j - \mathbf{x}_j^T\hat{\mathbf{t}}_k + \hat{d}_{j,k}\hat{c}_{\alpha_{j,k}},$$
$$[\hat{\mathbf{b}}_2]_{3N_{BS}+j-2J_S} = \mathbf{y}_j^T\mathbf{b}_j - \mathbf{y}_j^T\hat{\mathbf{t}}_k + \hat{d}_{j,k}\hat{c}_{\beta_{j,k}}, \quad (59)$$

$\hat{\mathbf{A}}_2$ is calculated by

$$[\hat{\mathbf{A}}_2]_{i_0,1\sim 6} = [2(\mathbf{b}_i - \hat{\mathbf{t}}_k)^T, \mathbf{0}_{1\times 3}],$$
$$[\hat{\mathbf{A}}_2]_{N_{BS}+i_0,1\sim 6} = [-2\hat{\mathbf{t}}_k, 2(\mathbf{b}_i - \hat{\mathbf{t}}_k)^T],$$
$$[\hat{\mathbf{A}}_2]_{2N_{BS}+j-J_S,1\sim 6} = [\mathbf{x}_j^T, \mathbf{0}_{1\times 3}],$$
$$[\hat{\mathbf{A}}_2]_{3N_{BS}+j-2J_S,1\sim 6} = [\mathbf{y}_j^T, \mathbf{0}_{1\times 3}], \quad (60)$$

And $\mathbf{T}_3$ is divided into four parts as $\mathbf{T}_3 \triangleq [\mathbf{T}_C^T, \mathbf{T}_D^T, \mathbf{T}_E^T, \mathbf{T}_F^T]^T \in \mathbb{R}^{(2N_{BS}+2J)\times(2N_{BS}+2J)}$, with $\mathbf{T}_C, \mathbf{T}_D \in \mathbb{R}^{(N_{BS}-1)\times(2N_{BS}+2J)}$, $\mathbf{T}_E, \mathbf{T}_F \in \mathbb{R}^{J\times(2N_{BS}+2J)}$. The non-zero elements are calculated via

$$[\mathbf{T}_C]_{i_0,i_0} = -2\hat{d}_{i_0,k},$$
$$[\mathbf{T}_D]_{i_0,i_0} = -2\hat{d}_{i_0,k}, [\mathbf{T}_D]_{i_0,N_{BS}+i_0} = -2\hat{d}_{i_0,k},$$
$$[\mathbf{T}_E]_{j-J_S,j} = \hat{c}_{\alpha_{j,k}}, [\mathbf{T}_E]_{j-J_S,2N_B+j-J_S} = \hat{d}_{j,k},$$
$$[\mathbf{T}_F]_{j-J_S,j} = \hat{c}_{\beta_{j,k}}, [\mathbf{T}_F]_{j-J_S,2N_B+J+j-J_S} = \hat{d}_{j,k}, \quad (61)$$

The second-stage WLS estimation of $\Delta\boldsymbol{\mu}_k$ is

$$\hat{\boldsymbol{\Theta}}_2 = (\hat{\mathbf{A}}_2^H\mathbf{W}_2\hat{\mathbf{A}}_2)^{-1}\hat{\mathbf{A}}_2^H\mathbf{W}_2\hat{\mathbf{b}}_2, \quad (62)$$

where

$$\mathbf{W}_2 = \mathbf{Q}_{\varepsilon_2}^{-1} = (\mathbf{T}_3^H\mathbf{Q}_{\boldsymbol{\mu}_k}\mathbf{T}_3)^{-1} = (\mathbf{T}_3^H(\mathbf{T}_1^H\mathbf{Q}_{m_k}^{-1}\mathbf{T}_1)^{-1}\mathbf{T}_3)^{-1}, \quad (63)$$

The covariance matrix of $\hat{\boldsymbol{\Theta}}_2$ is

$$\mathbf{Q}_{\boldsymbol{\Theta}_2} = (\hat{\mathbf{A}}_2^H\mathbf{W}_2\hat{\mathbf{A}}_2)^{-1}, \quad (64)$$

With two-stage WLS, the final estimation is $\hat{\boldsymbol{\Theta}}_{TS} = \hat{\boldsymbol{\Theta}}_1 + \hat{\boldsymbol{\Theta}}_2$.

*C. Closed-Form Fusion Algorithm for FD-NCS*

The DoF of range, or velocity, in FD-NCS equal to $N_{BS}$, while those of HD-NCS become $N_{BS}$-1. A simple proof for HD-NCS is that the $N_{BS}$ independent distances, or velocities cannot be recovered from measurements of $IJ$ pairs, while if one of $N_{BS}$ distances, or velocities, is added, the remaining parameters can be recovered. The dimension reduction operation of Equation (38) cannot be used. In this case, this paper proposes to use two nuisance parameters of $d_{0,k}$ and $\dot{d}_{0,k}$ to compensate the DoFs of distances and velocities, respectively. The following relationship is obtained as

$$\mathbf{m}_k - d_{0,k}\mathbf{v}_d - \dot{d}_{0,k}\mathbf{v}_{\dot{d}} = \mathbf{T'}_1\boldsymbol{\mu'}_k, \quad (65)$$

where $\mathbf{v}_d = [\mathbf{1}_J^T, \mathbf{0}_{(4IJ-J)}^T]^T$, $\mathbf{v}_{\dot{d}} = [\mathbf{0}_{IJ}^T, \mathbf{1}_J^T, \mathbf{0}_{3IJ-J}^T]^T$, $\mathbf{T'}_1 \in \mathbb{R}^{4IJ\times(2N_{BS}+2J-2)}$ is a matrix by removing the column 0 and $N_{BS}$ of $\mathbf{T}_1$, and $\boldsymbol{\mu'}_k$ is a vector by removing the element 0 and $N_{BS}$ of $\boldsymbol{\mu}_k$.

For this model, WLS is still used to compress the MM of all pairs, and

$$\hat{\boldsymbol{\mu}'}_k = (\mathbf{T'}_1^H\mathbf{W}_{m_k}\mathbf{T'}_1)^{-1}\mathbf{T'}_1^H\mathbf{W}_{m_k}(\hat{\mathbf{m}}_k - d_{0,k}\mathbf{v}_d - \dot{d}_{0,k}\mathbf{v}_{\dot{d}})$$
$$\triangleq \hat{\boldsymbol{\mu}''}_k - \mathbf{g}_d d_{0,k} - \mathbf{g}_{\dot{d}}\dot{d}_{0,k}. \quad (66)$$

The covariance matrix of $\hat{\boldsymbol{\mu}''}_k$ is therefore

$$\mathbb{E}[\Delta\boldsymbol{\mu''}_k\Delta\boldsymbol{\mu''}_k^T] = (\mathbf{T'}_1^H\mathbf{W}_{m_k}\mathbf{T'}_1)^{-1}, \quad (67)$$

Till this step, all elements in $\boldsymbol{\mu'}_k$ is expressed linearly to the nuisance parameters, $d_{0,k}$ and $\dot{d}_{0,k}$.

Using Equation (66), the distance in $\boldsymbol{\mu}_k$ is written as

$$d_{i_1,k} = [\boldsymbol{\mu''}_k]_{i_1-1} - [\mathbf{g}_d]_{i_1-1}d_{0,k} \triangleq d''_{i_1,k} - g_{i_1}d_{0,k}. \quad (68)$$

It can be deduced from the relationship between TX and RX BSs that $g_{i_1} = -1$ for the RX BSs with $i_1 = J_S,\ldots,J_S + J - 1$, and similarly, $g_{i_1} = 1$ for the other TX BSs with $i_1 = 1,\ldots,I-1$.

For the distance measurements, using Equation (68) and $|g_{i_1}| = 1$, Equation (43) can be manipulated algebraically into

$$2(\mathbf{b}_{i_1} - \mathbf{b}_0)^T\mathbf{t}_k - 2g_{i_1}d''_{i_1,k}d_{0,k} = \mathbf{b}_{i_1}^T\mathbf{b}_{i_1} - \mathbf{b}_0^T\mathbf{b}_0 - d''^2_{i_1,k}. \quad (69)$$

For the radial velocity measurements, Equation (44) is transformed into the following equation with $\dot{d}_{i_1,k} \triangleq \dot{d}''_{i_1,k} - g_{i_1}\dot{d}_{0,k}$ and $|g_{i_1}| = 1$.

$$2(\mathbf{b}_{i_1} - \mathbf{b}_0)^T\dot{\mathbf{t}}_k - 2g_{i_1}\dot{d}''_{i_1,k}d_{0,k} - 2g_{i_1}d''_{i_1}\dot{d}_{0,k} = -2d''_{i_1,k}\dot{d}''_{i_1,k}. \quad (70)$$

For the angle measurements, Equation (45) is rewritten into

$$\mathbf{x}_j^T\mathbf{t}_k - g_j\cos(\alpha_{j,k})d_{0,k} = \mathbf{x}_j^T\mathbf{b}_j - d''_{j,k}\cos(\alpha_{j,k}),$$
$$\mathbf{y}_j^T\mathbf{t}_k - g_j\cos(\beta_{j,k})d_{0,k} = \mathbf{y}_j^T\mathbf{b}_j - d''_{j,k}\cos(\beta_{j,k}), \quad (71)$$

As two nuisance parameters are introduced, the unknown variables become

$$\boldsymbol{\Theta'}_1 = [\boldsymbol{\Theta}_1, d_{0,k}, \dot{d}_{0,k}], \quad (72)$$

Stacking Equations (70)~(72) for $i_1 = 1, \ldots, N_{BS} - 1$ and $j = J_S, \ldots, J_S + J - 1$ and replacing true values with estimations and errors,

$$\boldsymbol{\varepsilon'}_1 = \mathbf{T'}_2 \Delta \boldsymbol{\mu''}_k = \widehat{\mathbf{b}}'_1 - \widehat{\mathbf{A}}'_1 \boldsymbol{\Theta'}_1, \tag{73}$$

where $\widehat{\mathbf{b}}_1$ is

$$\begin{aligned}
\left[\widehat{\mathbf{b}}_1\right]_{i_1-1} &= \mathbf{b}_{i_1}^T \mathbf{b}_{i_1} - \mathbf{b}_{i_1}^T \mathbf{b}_{i_1} - \widehat{d''}_{i_1,k}^2, \\
\left[\widehat{\mathbf{b}}_1\right]_{N_{BS}+i_1-2} &= -2\widehat{d''}_{i_1,k}\widehat{d''}_{i_1,k}, \\
\left[\widehat{\mathbf{b}}_1\right]_{2N_{BS}+j-J_S-2} &= \mathbf{x}_j^T \mathbf{b}_j + \widehat{d''}_{j,k}\hat{c}_{\alpha_{j,k}}, \\
\left[\widehat{\mathbf{b}}_1\right]_{3N_{BS}+j-2J_S-2} &= \mathbf{y}_j^T \mathbf{b}_j + \widehat{d''}_{j,k}\hat{c}_{\beta_{j,k}}, \tag{74}
\end{aligned}$$

$\widehat{\mathbf{A}}'_1$ is calculated via

$$\begin{aligned}
\left[\widehat{\mathbf{A}}'_1\right]_{i_1-1,1\sim 8} &= \left[2(\mathbf{b}_{i_1} - \mathbf{b}_0)^T, \mathbf{0}_3^T, -2g_{i_1}\widehat{d''}_{i_1,k}, 0\right], \\
\left[\widehat{\mathbf{A}}'_1\right]_{N_{BS}+i_1-2,1\sim 8} &= \left[\mathbf{0}_3^T, 2(\mathbf{b}_i - \mathbf{b}_0)^T, -2g_{i_1}\widehat{d''}_{i_1,k}, -2g_{i_1}\widehat{d''}_{i_1,k}\right], \\
[\mathbf{A}_1]_{2N_{BS}+j-J_S-2,1\sim 8} &= [\mathbf{x}_j^T, \mathbf{0}_3^T, g_j\cos(\alpha_{j,k}), 0], \\
[\mathbf{A}_1]_{3N_{BS}+j-2J_S-2,1\sim 8} &= [\mathbf{y}_j^T, \mathbf{0}_3^T, g_j\cos(\beta_{j,k}), 0], \tag{75}
\end{aligned}$$

and $\mathbf{T'}_2 \triangleq \left[\mathbf{T'}_C^T, \mathbf{T'}_D^T, \mathbf{T'}_E^T, \mathbf{T'}_F^T\right]^T \in \mathbb{R}^{(2N_{BS}+2J-2)\times(2N_{BS}+2J-2)}$, where the non-zero elements in $\mathbf{T'}_C, \mathbf{T'}_D \in \mathbb{R}^{(N_{BS}-1)\times(2N_{BS}+2J)}$ and $\mathbf{T'}_E, \mathbf{T'}_F \in \mathbb{R}^{J\times(2N_{BS}+2J)}$ are

$$\begin{aligned}
[\mathbf{T'}_C]_{i_1-1,i_1-1} &= -2\widehat{d''}_{i_1,k}, \\
[\mathbf{T'}_D]_{i_1-1,i_1-1} &= -2\widehat{d''}_{i_1,k}, [\mathbf{T'}_D]_{i_1-1,N_{BS}-2+i_1} = -2\widehat{d''}_{i_1,k}, \\
[\mathbf{T'}_E]_{j-J_S,j-1} &= \hat{c}_{\alpha_{j,k}}, [\mathbf{T'}_E]_{j-J_S,2N_{BS}-2+j-J_S} = \widehat{d''}_{j,k}, \\
[\mathbf{T'}_F]_{j-J_S,j-1} &= \hat{c}_{\beta_{j,k}}, [\mathbf{T'}_F]_{j-J_S,2N_{BS}+J-2+j-J_S} = \widehat{d''}_{j,k}, \tag{76}
\end{aligned}$$

The first-stage WLS can be done to obtain the estimations of the positions, velocities and nuisance parameters,

$$\widehat{\boldsymbol{\Theta}}'_1 = \left(\widehat{\mathbf{A}}_1^{\prime H} \mathbf{W'}_1 \widehat{\mathbf{A}}'_1\right)^{-1} \widehat{\mathbf{A}}_1^{\prime H} \mathbf{W'}_1 \widehat{\mathbf{b}}'_1, \tag{77}$$

where

$$\mathbf{W'}_1 = \mathbf{Q}_{\boldsymbol{\varepsilon'}_1}^{-1} = \left(\mathbf{T'}_2^H \mathbf{Q}_{\boldsymbol{\mu''}_k} \mathbf{T'}_2\right)^{-1} = \left(\mathbf{T'}_2^H \left(\mathbf{T'}_1^H \mathbf{Q}_{\mathbf{m}_k}^{-1} \mathbf{T'}_1\right)^{-1} \mathbf{T'}_2\right)^{-1}, \tag{78}$$

The covariance matrix of $\widehat{\boldsymbol{\Theta}}'_1$ is obtained as

$$\mathbf{Q}_{\boldsymbol{\Theta'}_1} = \left(\widehat{\mathbf{A}}_1^{\prime H} \mathbf{W'}_1 \widehat{\mathbf{A}}'_1\right)^{-1}, \tag{79}$$

For the first-stage WLS here, the number of equations equals to the DoF of measurements. However, two nuisance parameters are used, which is dependent on the desired $\boldsymbol{\mu}_k$. Therefore, the results of first-stage WLS is not the optimal, and the second stage refinement is required. The second-stage WLS is still based on Equation (58). The first challenge is calculating the covariance matrix of $\boldsymbol{\mu}_k$. The $d_{0,k}$ and $\dot{d}_{0,k}$ in $\boldsymbol{\mu}_k$ is linear combinations of elements in $\boldsymbol{\mu''}_k$ as

$$\Delta d_{0,k} = \mathbf{g}_7^T \Delta \boldsymbol{\mu''}_k, \Delta \dot{d}_{0,k} = \mathbf{g}_8^T \Delta \boldsymbol{\mu''}_k, \tag{80}$$

where $\mathbf{g}_7 \triangleq [\mathbf{G}]_{7,1\sim 2N_{BS}+2J-2}^T$, $\mathbf{g}_8 \triangleq [\mathbf{G}]_{8,1\sim 2N_{BS}+2J-2}^T$, and $\mathbf{G} \triangleq \left(\widehat{\mathbf{A}}_1^{\prime H} \mathbf{W'}_1 \widehat{\mathbf{A}}'_1\right)^{-1} \widehat{\mathbf{A}}_1^{\prime H} \mathbf{W'}_1 \mathbf{T'}_2^H$.

With Equation (80), $\Delta \boldsymbol{\mu'}_k$ is obtained using linear transformation of $\Delta \boldsymbol{\mu''}_k$ as

$$\begin{aligned}
\Delta \boldsymbol{\mu'}_k &= \Delta \boldsymbol{\mu''}_k - \mathbf{g}_d \mathbf{g}_7^T \Delta \boldsymbol{\mu''}_k - \mathbf{g}_d \mathbf{g}_8^T \Delta \boldsymbol{\mu''}_k \\
&= \left(\mathbf{I}_{2N_{BS}+2J-2} - \mathbf{g}_d \mathbf{g}_7^T - \mathbf{g}_d \mathbf{g}_8^T\right) \Delta \boldsymbol{\mu''}_k, \tag{81}
\end{aligned}$$

Combining the results in Equations (80) and (81), $\Delta \boldsymbol{\mu}_k$ is obtained via $\Delta \boldsymbol{\mu}_k = \mathbf{T'}_3 \Delta \boldsymbol{\mu''}_k$, where $[\mathbf{T'}_3]_{1,1\sim N_{eq}} = \mathbf{g}_7^T$, $[\mathbf{T'}_3]_{N_{BS}+1,1\sim N_{eq}} = \mathbf{g}_8^T$, and $[\mathbf{T'}_3]_{[2\sim N_{BS},N_{BS}+1\sim N_{eq}],1\sim N_{eq}} = \mathbf{I}_{2N_{BS}+2J-2} - \mathbf{g}_d \mathbf{g}_7^T - \mathbf{g}_d \mathbf{g}_8^T$.

Another challenge is that $\mathbf{W}_2 = \left(\mathbf{T}_3^H \mathbf{T}_3^{\prime H} \mathbf{Q}_{\boldsymbol{\mu''}_k}^{-1} \mathbf{T'}_3 \mathbf{T}_3\right)^{-1}$ cannot be calculated as $\mathbf{T'}_3^H \mathbf{Q}_{\boldsymbol{\mu''}_k}^{-1} \mathbf{T'}_3$ is rank-deficient. To solve this problem, a random sampling matrix $\mathbf{S} \in \mathbb{C}^{(2N_{BS}+2J-2)\times(2N_{BS}+2J)}$ is employed to linearly transform Equation (58) into

$$\boldsymbol{\varepsilon'}_2 = \mathbf{S}\mathbf{T}_3 \Delta \boldsymbol{\mu}_k = \mathbf{S}\widehat{\mathbf{b}}_2 - \mathbf{S}\widehat{\mathbf{A}}_2 \boldsymbol{\Theta}_2, \tag{82}$$

The second-stage WLS estimation of $\Delta \boldsymbol{\mu}_k$ becomes

$$\widehat{\boldsymbol{\Theta}}'_2 = \left(\widehat{\mathbf{A}}_2^H \mathbf{S}^H \mathbf{W'}_2 \mathbf{S}\widehat{\mathbf{A}}_2\right)^{-1} \widehat{\mathbf{A}}_2^H \mathbf{S}^H \mathbf{W'}_2 \mathbf{S}\widehat{\mathbf{b}}_2, \tag{83}$$

where

$$\mathbf{W'}_2 = \mathbf{Q}_{\boldsymbol{\varepsilon'}_2}^{-1} = \left(\mathbf{S}^H \mathbf{T}_3^H \mathbf{T}_3^{\prime H} \left(\mathbf{T}_1^{\prime H} \mathbf{Q}_{\mathbf{m}_k}^{-1} \mathbf{T'}_1\right)^{-1} \mathbf{T'}_3 \mathbf{T}_3 \mathbf{S}\right)^{-1}, \tag{84}$$

The covariance matrix of $\widehat{\boldsymbol{\Theta}}'_2$ is

$$\mathbf{Q}_{\boldsymbol{\Theta'}_2} = \left(\widehat{\mathbf{A}}_2^H \mathbf{S}^H \mathbf{W'}_2 \mathbf{S}\widehat{\mathbf{A}}_2\right)^{-1}, \tag{85}$$

The final estimation is $\widehat{\boldsymbol{\Theta}}'_{TS} = \left[\widehat{\boldsymbol{\Theta}}'_1\right]_{1\sim 6} + \widehat{\boldsymbol{\Theta}}'_2$.

## VI. NUMERICAL RESULTS

In this section, simulation results are shown to verify the proposed schemes. The simulation parameters based on 3GPP protocol [43] are listed in Table I. Here, a sub-6G BS is assumed, which is widely used in current networks. The maximum bandwidth configuration in sub-6G bandwidth is 100 MHz. Considering guard bands, the effective bandwidth is 98.28 MHz. The TX BSs transmit one OFDM symbol every 1 ms, and the processing is done every 64 ms. The positions and orientations of BSs are listed in Table II. In practice, $\mathbf{x}_j$ is parallel to the ground, which means $[\mathbf{x}_j]_1 = 0$. As $\mathbf{x}_j$, $\mathbf{y}_j$ and $\mathbf{z}_j$ are orthogonal, $\mathbf{x}_j^T \mathbf{y}_j = \mathbf{x}_j^T \mathbf{z}_j = \mathbf{y}_j^T \mathbf{z}_j = 0$. By taking $\mathbf{z}_j$ as a known variable, these equations can be easily solved as

$$\mathbf{x}_j = \frac{\left[-[\mathbf{z}_j]_2, [\mathbf{z}_j]_1, 0\right]^T}{\left\|\left[-[\mathbf{z}_j]_2, [\mathbf{z}_j]_1, 0\right]^T\right\|}, \mathbf{y}_j = \frac{\left[-[\mathbf{z}_j]_1[\mathbf{z}_j]_3, -[\mathbf{z}_j]_2[\mathbf{z}_j]_3, [\mathbf{z}_j]_1^2+[\mathbf{z}_j]_2^2\right]^T}{\left\|\left[-[\mathbf{z}_j]_1[\mathbf{z}_j]_3, -[\mathbf{z}_j]_2[\mathbf{z}_j]_3, [\mathbf{z}_j]_1^2+[\mathbf{z}_j]_2^2\right]^T\right\|}. \quad (86)$$

TABLE I SIMULATION PARAMETERS

| Parameter | Value |
|---|---|
| Carrier frequency, $f_C$ | 4.9 GHz |
| Sub-carrier spacing, $\Delta f$ | 30 kHz |
| Total sub-carrier number, $N$ | 3276 |
| Effective bandwidth, $B$ | 98.28 MHz |
| Noise power density | –174 dBm/Hz |
| Noise figure | 6 dB |
| Pulse repetition interval, $T$ | 1 ms |
| OFDM symbol number, $M$ | 64 |
| UPA size, $(L_x, L_y)$ | (8, 8) |
| Radar cross section | 1 m$^2$ |
| BS number, $N_{BS}$ | 4 for FD-NCS & 5 for HD-NCS |
| TX BS number, $I$ | 2 for FD-NCS & 3 for HD-NCS |
| Target number, $K$ | 3 |

TABLE II BS POSITIONS AND ORIENTATIONS

|  | Position (m) | Orientation Vector, $\mathbf{z}_j$ |
|---|---|---|
| BS 0 | (0, 0, 80) | (0.7032, 0.7032, -0.1045) |
| BS 1 | (500, 0, 20) | (0.7032, -0.7032, -0.1045) |
| BS 2 | (0, 500, 20) | (-0.7032, 0.7032, -0.1045) |
| BS 3 | (500, 500, 80) | (-0.7032, -0.7032, -0.1045) |
| BS 4 | (250, 500, 40) | (0, -0.9945, -0.1045) |

TABLE III TARGET POSITIONS AND VELOCITY

|  | Target 0 | Target 1 | Target 2 |
|---|---|---|---|
| Position (m) | (125, 250, 0) | (250, 250, 60) | (375, 250, 30) |
| Velocity (m/s) | (10, 10, 0) | (10, -5, -5) | (-5, -5, -5) |

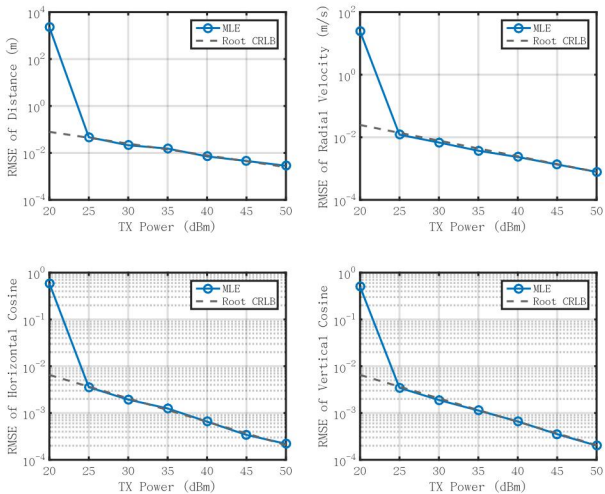

**Fig. 5.** The RMSE vs root CRLB of MM in a one-target case.

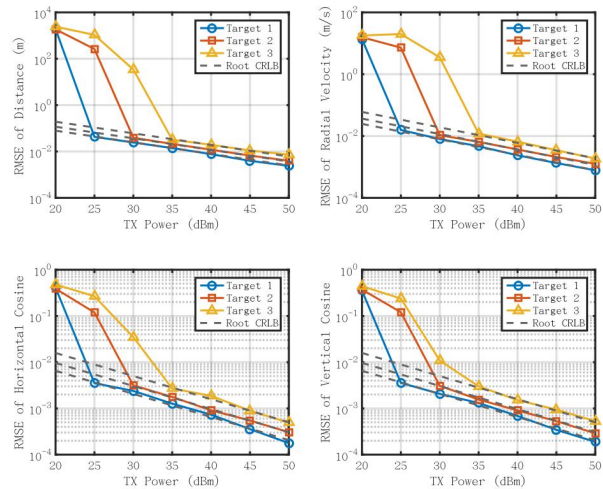

**Fig. 6.** The RMSE vs root CRLB of MM in a three-target case using the proposed AD-NOMP.

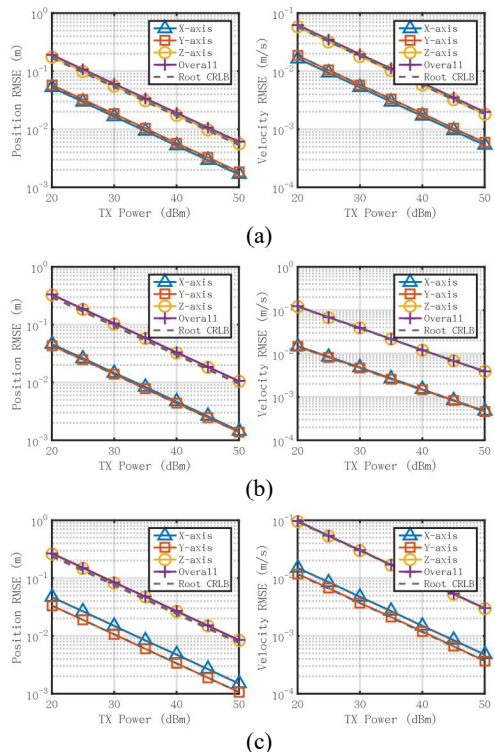

**Fig. 7.** The position and velocity RMSE of (a) Target 0, (b) Target 1, and (c) Target 2 using ideal MM estimations and DoF-TSWLS in the FD-NCS scenario.

Althrough the DoF analysis in the previous section, the theoretical minimum $N_{BS}$ to jointly recover the 3D position and velocity is 3 and 4 in FD-NCS and HD-NCS. The proposed DoF-TSWLS only requires one more BS in both cases.

In Fig. 5, a one-target scenario using target 0 is simulated, and the root mean square error (RMSE) of MM is compared with the root CRLB. This one-target simulation is used to verify the CRLB of MM and proposition 1. The MLE here is realized by the AD-NOMP, as it is an MLE when there is only one target. The simulation results show that the CRLB of target 0 can be reached with a TX power of not less than 25 dBm in the (0,0) TX-RX pairs. After the verification of the CRLB of MM, the proposed AD-NOMP is verified in the three-target case in Fig. 6. This result shows that the MM CRLBs of multiple targets are reached with a certain signal power to noise ratio (SNR). It also implies that the joint multi-domain processing of AD-NOMP can effectively utilizes multiple domians to suppress the multi-target interference.

Fig. 7 shows the position and velocity RMSE of multiple targets using ideal MM estimations and the proposed DoF-TSWLS in the FD-NCS scenario. The ideal MM estimations is directly generated with errors reaching CRLB of MM. The use of ideal MM estimations is to independently verify the performance of the fusion algorithm. The simulated results show that the RMSE of DoF-TSWLS is tightly bounded to the corresponding CRLB. Also, it can be found that the position and velocity in the Z-axis, or the height dimension, have the lowest accuracy. This is due to the fact that the height differences of multiple BSs are relatively small, which makes the Z-axis estimation more sensitive to noise. With this feature, the RMSE at Z-axis also dominates the overall RMSE of 3D position and velocity.

Fig. 8 shows the position and velocity RMSE of multiple targets using ideal MM estimations and DoF-TSWLS in the

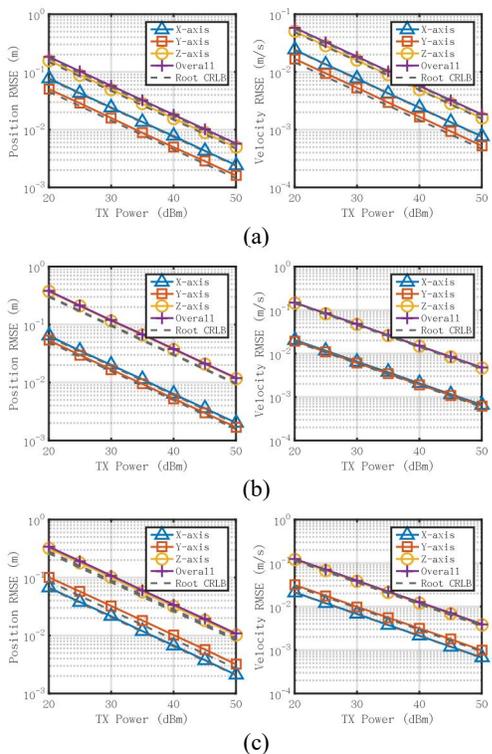

**Fig. 8.** The position and velocity RMSE of (a) Target 0, (b) Target 1, and (c) Target 2 using ideal MM estimations and DoF-TSWLS in the HD-NCS scenario.

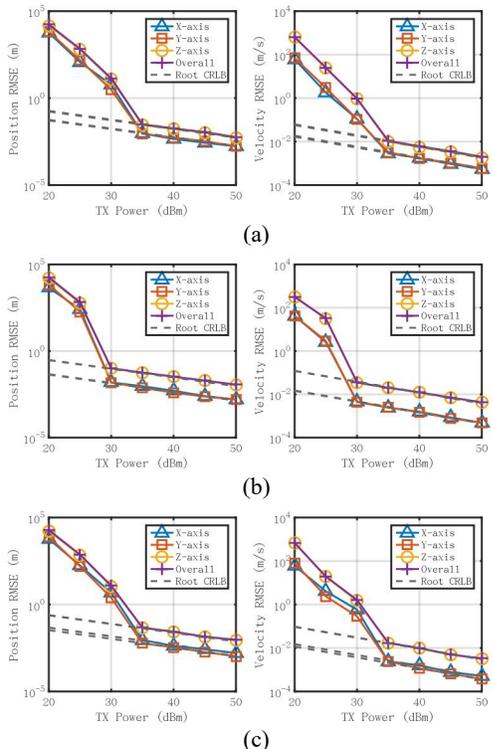

**Fig. 9.** The position and velocity RMSE of (a) Target 0, (b) Target 1, and (c) Target 2 using AD-NOMP estimations and DoF-TSWLS in the FD-NCS scenario.

HD-NCS scenario. The RMSE is close to the corresponding CRLB, but there are small gaps between the RMSE and the root CRLB for the position and velocity estimations of some targets in some axes. It may result from introducing nuisance parameters, which leads to some extra processing operations. Omitting higher-order errors in these operations brings about larger errors than those of FD-NCS. Similarly, the position and velocity in the Z-axis have higher estimation errors than those in the other axes.

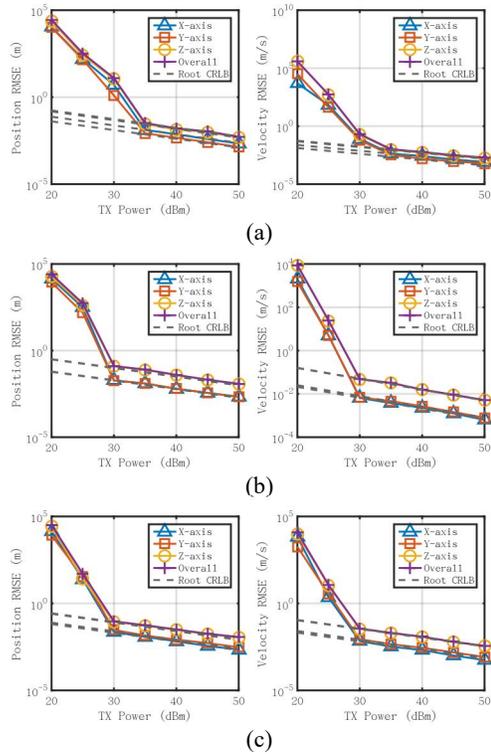

**Fig. 10.** The position and velocity RMSE of (a) Target 0, (b) Target 1, and (c) Target 2 using AD-NOMP estimations and DoF-TSWLS in the HD-NCS scenario.

The position and velocity estimation RMSE performances using AD-NOMP estimations are shown in Fig. 9 and Fig. 10. As enough SNR is required to attain the CRLB of MM in AD-NOMP, the RMSE of joint AD-NOMP and DoF-TSWLS also requires certain SNRs to approach CRLB of positions and velocities. Apart from the requirement of enough SNR to approach the CRLB, the results are very similar to those of Fig. 7 and 8. These results also show that the quality of MM estimations should be identified in the fusion steps to avoid large fusion errors from inaccurate MM.

The derived CRLB of position and velocity in this paper is determined by the system parameters, instead of a known covariance matrix. It can be used to instruct the NCS system design. In Fig. 11, the performance limits versus $N_{BS}$ are shown. In this simulation, $P_{TX} = 35$ dBm, $I = 1$, and $J$ is therefore $N_{BS}$ and $N_{BS} - 1$ in FD-NCS and HD-NCS. As mentioned before, the Z-axis position and velocity are more difficult to be accurately estimated. Therefore, the 2D performance limits are much better than those of 3D. Also, in both FD-NCS and FD-NCS, the performance limits of 2D positioning decrease very fast when $N_{BS}$ is small. This is caused by that the positioning still relies on the angle estimation for a small $N_{BS}$. With the increase of $N_{BS}$, the performance improvement becomes much slower, as these performance limits are only dependent on the range measurements. More BSs are required for the 3D positioning to gain the turning point of performance limits due to the unstable Z-axis performance. For the velocity estimations, there are minimum BS requirements of 3 and 4 in FD-NCS and HD-NCS, respectively. The performance limits of these two cases are very close.

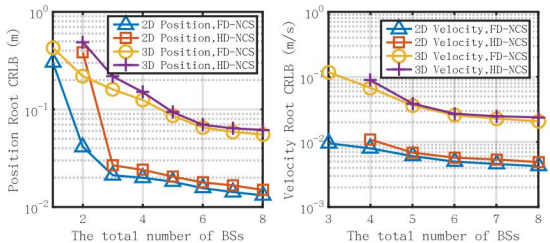

**Fig. 11.** The performance limits versus the total number of BSs.

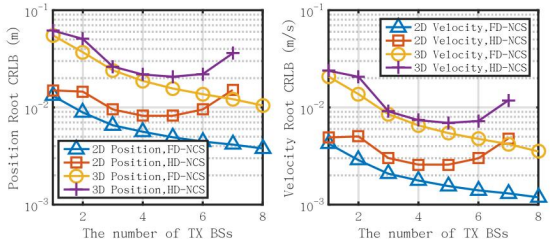

**Fig. 12.** The performance limits versus the number of TX BSs.

Fig. 12 shows the performance limits with different $I$. In this simulation, $P_{TX} = 35$ dBm, $N_{BS} = 8$, and $J$ is therefore $N_{BS}$ and $N_{BS} - I$ in FD-NCS and HD-NCS, respectively. For the position and velocity estimations of FD-NCS, the increment of $I$ always makes the performance limits lower. This implies that $I$ should be as large as possible to achieve a better performance in the FD-NCS. For those of the HD-NCS, there is an optimal value of $I$ to obtain best performance limits. That is to say, given a group of NCS BSs, the TX and RX BSs should be carefully arranged to optimize the sensing performance.

## VII. CONCLUSION

This paper proposes and researches NCS-MM in terms of modelling, protocol, performance limits and algorithms under the settings of cellular networks. It proposes a novel multi-domain decoupling model to simplify NCS-MM and a GP-based protocol to realize highly efficient NCS-MM. With the multi-domain decoupling model, the CRLB of MM can be derived from a simple single-dimension accumulated channel, and the equivalence to the multi-dimension channel is also proven. Then, the CRLB of target position and velocity is derived using the CRLB of MM. Using the decoupling model, AD-NOMP is also proposed, which extends NOMP to arbitrary dimensions. Closed-form DoF-TSWLS algorithms are proposed for both FD-NCS and HD-NCS, which reduces the number of equations according to the DoF of MM and still gains good performance close to performance limits. In the future, both FD and HD NCS systems of higher estimation accuracy, low SNR adaptability, flexible BS switching, more targets and less BSs are recommended to be further researched.

## APPENDIX A
### THE PROOF OF PROPOSITION 1

In this appendix, proposition 1 is proven. Using Equation (15), it is easy to have

$$[\mathbf{J}(\boldsymbol{\Psi}_{i,j,k})]_{0,1\sim5}^{\mathrm{T}} = [\mathbf{J}(\boldsymbol{\Psi}_{i,j,k})]_{1\sim5,0} = \mathbf{0}_5, \quad (87)$$

Then, $\mathbf{J}(\boldsymbol{\Psi}_{i,j,k})^{-1}$ can be simplified into

$$\mathbf{J}^{-1}(\boldsymbol{\Psi}_{i,j,k}) = \begin{bmatrix} [\mathbf{J}(\boldsymbol{\Psi}_{i,j,k})]_{1,1}^{-1} & \mathbf{0}_5^{\mathrm{T}} \\ \mathbf{0}_5 & \left([\mathbf{J}(\boldsymbol{\Psi}_{i,j,k})]_{2\sim5,2\sim5}\right)^{-1} \end{bmatrix}, (88)$$

The $[\mathbf{J}(\boldsymbol{\Psi}_{i,j,k})]_{1\sim5,1\sim5}$ is divided into 4 parts as

$$[\mathbf{J}(\boldsymbol{\Psi}_{i,j,k})]_{1\sim5,1\sim5} = \begin{bmatrix} [\mathbf{J}(\boldsymbol{\Psi}_{i,j,k})]_{1,1} & [\mathbf{J}(\boldsymbol{\Psi}_{i,j,k})]_{1,2\sim5} \\ [\mathbf{J}(\boldsymbol{\Psi}_{i,j,k})]_{2\sim5,1} & [\mathbf{J}(\boldsymbol{\Psi}_{i,j,k})]_{2\sim5,2\sim5} \end{bmatrix}, (89)$$

For Equation (89), the block inverse formula [44] can be used, and the formula is written as

$$\begin{bmatrix} \mathbf{A} & \mathbf{B} \\ \mathbf{C} & \mathbf{D} \end{bmatrix}^{-1} = \begin{bmatrix} \mathbf{A}^{-1} + \mathbf{A}^{-1}\mathbf{B}\mathbf{E}^{-1}\mathbf{C}\mathbf{A}^{-1} & -\mathbf{A}^{-1}\mathbf{B}\mathbf{E}^{-1} \\ -\mathbf{E}^{-1}\mathbf{C}\mathbf{A}^{-1} & \mathbf{E}^{-1} \end{bmatrix}, (90)$$

where

$$\mathbf{E} = \mathbf{D} - \mathbf{C}\mathbf{A}^{-1}\mathbf{B}. \quad (91)$$

Only $\mathbf{E}^{-1} \triangleq [\mathbf{J}^{-1}(\boldsymbol{\Psi}_{i,j,k})]_{4,4}$ affects the CRLB of MM, which can be calculated by

$$\mathbf{E} = [\mathbf{J}(\boldsymbol{\Psi}_{i,j,k})]_{2\sim5,2\sim5} - [\mathbf{J}(\boldsymbol{\Psi}_{i,j,k})]_{2\sim5,1}[\mathbf{J}(\boldsymbol{\Psi}_{i,j,k})]_{1,1}^{-1}[\mathbf{J}(\boldsymbol{\Psi}_{i,j,k})]_{2\sim5,1}. \quad (92)$$

Using Equation (18), the non-diagonal elements $\mathbf{E}_{i',j'}$, where $i' \neq j'$, can be calculated by

$$\mathbf{E}_{i',j'} = [\mathbf{J}(\boldsymbol{\Psi}_{i,j,k})]_{i'+2,j'+2} - \frac{[\mathbf{J}(\boldsymbol{\Psi}_{i,j,k})]_{i'+2,1}[\mathbf{J}(\boldsymbol{\Psi}_{i,j,k})]_{j'+2,1}}{[\mathbf{J}(\boldsymbol{\Psi}_{i,j,k})]_{1,1}}$$
$$= \frac{2}{\sigma_w^2} A_{i,j,k}^2 \frac{N_0 N_1 N_2 N_3}{N_{i'}N_{j'}} \sum_{n_{i'}} 2\pi n_{i'} \sum_{n_{j'}} 2\pi n_{j'}$$
$$- \frac{\left(\frac{2}{\sigma_w^2} A_{i,j,k}^2 \frac{N_0 N_1 N_2 N_3}{N_{i'}} \sum_{n_{i'}} 2\pi n_{i'}\right)\left(\frac{2}{\sigma_w^2} A_{i,j,k}^2 \frac{N_0 N_1 N_2 N_3}{N_{j'}} \sum_{n_{j'}} 2\pi n_{j'}\right)}{\frac{2}{\sigma_w^2} A_{i,j,k}^2 N_0 N_1 N_2 N_3} = 0, (93)$$

Equation (93) shows that $\mathbf{E}$ is diagonal. The diagonal elements are calculated by

$$\mathbf{E}_{i',i'} = \frac{2}{\sigma_w^2} A_{i,j,k}^2 \frac{N_0 N_1 N_2 N_3}{N_{i'}} \sum_{n_{i'}} (2\pi n_{i'})^2$$
$$- \left(\frac{2}{\sigma_w^2} A_{i,j,k}^2 \frac{N_0 N_1 N_2 N_3}{N_{i'}} \sum_{n_{i'}} 2\pi n_{i'}\right) \bigg/ \frac{2}{\sigma_w^2} A_{i,j,k}^2 N_0 N_1 N_2 N_3$$
$$= \frac{2}{\sigma_w^2} A_{i,j,k}^2 \frac{N_0 N_1 N_2 N_3}{N_{i'}} \left(\frac{N_{i'}(N_{i'}-1)(2N_{i'}-1)}{6} - \frac{N_{i'}^2(N_{i'}-1)^2}{4N_{i'}}\right)$$
$$= \frac{A_{i,j,k}^2 N_0 N_1 N_2 N_3 (N_{i'}^2-1)}{6\sigma_w^2} = \frac{1}{J_{i,j,k}^{\mathrm{DM}i'}}. \quad (94)$$

Therefore,

$$[\mathbf{J}^{-1}(\boldsymbol{\Psi}_{i,j,k})]_{4\times4} = \mathbf{E}^{-1} = \mathrm{diag}(J_{i,j,k}^{\mathrm{DM0}}, J_{i,j,k}^{\mathrm{DM1}}, J_{i,j,k}^{\mathrm{DM2}}, J_{i,j,k}^{\mathrm{DM3}}), (95)$$

The proposition 1 is proven.

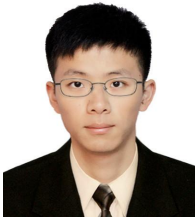
**Yihua Ma** (M'19) received the B.E. degree in information engineering from Southeast University, Nanjing, China in 2015 and the M.S. degree in microelectronics and solid state electronics from Peking University, Beijing, China in 2018.

Since 2018, he is with ZTE Corporation. He is now a pre-research senior expert in the Department of Wireless Algorithm, ZTE, Beijing, China and also a member of State Key Laboratory of Mobile Network and Mobile Multimedia Technology, Shenzhen, China. He is the first author of more than 10 articles, and more than 30 inventions. His main research interests include integrated sensing and communications, physical-layer design, grant-free transmissions, and massive MIMO. He is the project manager of several ZTE Industry-University-Institute Cooperation Funds.

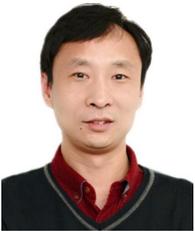
**Shuqiang Xia** received the B.S. degree in electronic engineering from Xiangtan University, Xiangtan, China, in 1999 and the M.S. degree in signal and information processing from Nanjing University of Science and Technology, China, in 2002.

He is a senior communication research expert in ZTE. His research interests focus on Integrated Sensing and Communications (ISAC), carrier aggregation (CA), and Ultra-Reliable Low-Latency Communications (URLLC).

Mr. Xia was a recipient of the China Patent Gold Award and Second Prize of National Technological Invention Award.

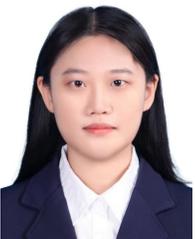
**Chen Bai** received the B.S. degree in Information Engineering from Nanjing University of Aeronautics and Astronautics, Nanjing, China, in 2019 and the M.S. degree in Electronics and Communications Engineering from Harbin Engineering University, Harbin, China, in 2022.

From 2022 to 2024, she was a technical pre-research engineer of ZTE Corporation, Shenzhen, China. Her research interests includes Integrated Sensing and Communication, adaptive control, and underwater acoustic signal processing..

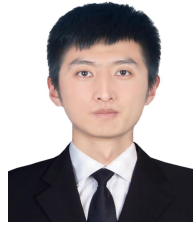
**Zhongbin Wang** received the B.S. degree in communication engineering from Chongqing University, Chongqing, China, in 2017 and Ph.D. degree in signal and information processing from University of the Chinese Academy of Sciences in 2022.

He is currently an engineer in ZTE Corporation. His research interests include the theory and application of Intergrated Sensing and Communications (ISAC), object detection, and environment sensing.

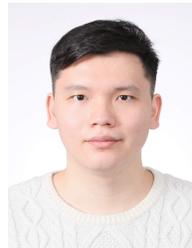
**Songqian Li** received the B.S. degree in electrical and computer engineering in 2020 and the M.S. degree in information and communication engineering in 2023 from Shanghai Jiao Tong University, Shanghai, China.

He joined ZTE corporation as a pre-research engineer in 2023. His research interests include intergrated sensing and communications, mmWave communications and backscatter communications.